\documentclass[%
 reprint,
 amsmath,amssymb,
 aps,
]{revtex4-2}

\usepackage{graphicx}% Include figure files
\usepackage{dcolumn}% Align table columns on decimal point
\usepackage{bm}% bold math
\usepackage{upgreek}
\newcommand{\vecsym}[1]{\boldsymbol{\mathrm{#1}}}
\usepackage{cleveref}   
\usepackage{color}
\usepackage{subcaption}
\usepackage{caption}
\begin{document}

% \preprint{APS/123-QED}

\title{Fluid-mediated impact of soft solids}% Force line breaks with \\
%\thanks{A footnote to the article title}%

\author{Jacopo Bilotto}
    \altaffiliation{Institute of Civil Engineering,
            \'{E}cole Polytechnique F\'{e}d\'{e}rale de Lausanne (EPFL), CH 1015 Lausanne, 
            Switzerland}
\author{John M. Kolinski}
     \altaffiliation{Institute of Mechanical Engineering,
            \'{E}cole Polytechnique F\'{e}d\'{e}rale de Lausanne (EPFL), CH 1015 Lausanne, 
            Switzerland}
\author{Brice Lecampion}
            \altaffiliation{Institute of Civil Engineering,
            \'{E}cole Polytechnique F\'{e}d\'{e}rale de Lausanne (EPFL), CH 1015 Lausanne, 
            Switzerland}
\author{Jean-François Molinari}
            \altaffiliation{Institute of Civil Engineering,
            \'{E}cole Polytechnique F\'{e}d\'{e}rale de Lausanne (EPFL), CH 1015 Lausanne, 
            Switzerland}
\author{Ghatu Subhash}
            \altaffiliation{Department of Mechanical and Aerospace Engineering,\\ University of Florida, Gainesville, FL, 32611, United States of America}
\author{Joaquin Garcia-Suarez}
            \altaffiliation{Institute of Civil Engineering,
            \'{E}cole Polytechnique F\'{e}d\'{e}rale de Lausanne (EPFL), CH 1015 Lausanne, 
            Switzerland}

\date{\today}% It is always \today, today,
             %  but any date may be explicitly specified

\begin{abstract}
A viscous, lubrication-like response can be triggered in a thin film of fluid squeezed between a rigid and flat surface and the tip of an incoming projectile. 
We develop a comprehensive theory for this viscous approach stage of fluid-mediated normal impact, applicable to soft impactors. 
Under the assumption of mediating fluid being incompressible, the impacting solid displays two limit regimes: one dominated by elasticity and the other by inertia.
The transition between the two is predicted by a dimensionless parameter, which can be interpreted as the ratio between two time scales that are
the time that it takes for the surface waves to warn the leading edge of the impactor of the forthcoming impact, 
and the characteristic duration of the final viscous phase of the approach. 
Additionally, we assess the role of solid compressibility and elucidate why nearly-incompressible solids feature 
(a) substantial ``gliding'' prior to contact at the transition between regimes, 
(b) the largest size of entrapped bubble between the deformed tip of the impactor and the flat surface, 
and (c) a sudden drop in entrapped bubble radius past the transition between regimes.    
Finally, we argue that the above time scale ratio (a dimensionless number) can govern the different dynamics reported experimentally for a fluid droplet as a function of its viscosity and surface tension. 
\end{abstract}
%\keywords{Suggested keywords}%Use showkeys class option if keyword
                              %display desired
\maketitle

%\tableofcontents
\section{Introduction}

Impacts between solid bodies occur daily in many natural and industrial processes. 
For stiff solids, the role of ambient air is negligible; however, compliant solids can be deformed by the lubrication stress in an intervening fluid layer \textcolor{black}{\cite{Rallabandi:2023}}. 
In this scenario, the lubricant can become entrained, and depending on the contact velocity, the contact area can be reduced \cite{Zheng:2021}. 
Fluid mediated contact between soft bodies governs myriad biological and physical processes, from the adhesion of hydrogel tapes \cite{xue:2021}, to the hemodynamics of blood cells in the vasculature \cite{higgins:2010} and the rheology of suspensions \cite{coussot:1999}.
Even the mundane act of holding a dish in the kitchen with our fingers is mediated by the surrounding air.
Despite the ubiquity and importance of fluid mediated contact between soft solids, surprisingly little is known about the dynamics of contact formation; 
indeed, for m/sec impacts and solids with a shear modulus similar to biological soft tissues, an observed sharp transition of the entrained air suggests a dominant role of elastodynamics, as the deformation rates in the solid transition from super-Rayleigh to sub-Rayleigh speeds \cite{Zheng:2021}.

Contact formation between compliant solids in air is analogous to liquid droplet impact; in this context, the lubricating air film generates the first formation of an on-axis dimple before ultimately leading to droplet splashing or spreading \cite{Mandre:2009, Mandre:2010, kolinski:2012,Thoroddsen:2016, Wu:2021, riboux:2014}.
Theoretical models of ``\textit{air cushioning}'' have been developed \cite{smith:2003, Hicks:2010} and validated \cite{Purvis:2012} for low viscosity droplets.
It has been reported that the otherwise reasonable, and usually made, assumptions of incompressibility and negligible surface tension break down, and even the air can become rarefied \cite{Mandre:2010, hicks+purvis:2013}. 
Remarkably, if these extra effects did not enter the picture, normal contact would be mathematically impossible \cite{Lin:2001, hillairet:2007}.
Experiments of highly viscous droplets impacting rigid \cite{langley:2017} and fluid surfaces \cite{langley:2019} displayed ``\textit{gliding}'' (local hovering around the tip of the droplet's leading edge while the outward regions continue approaching), 
whereas droplets impacting on a compliant surface entrain a larger amount of air compared to the rigid case \cite{langley:2020}.
Interestingly, a maximum in air bubble volume has been observed as a function of surface tension \cite{Bouwhuis:2012}.

\begin{figure}
    \centering
    %\captionsetup[subfigure]{justification=RaggedLeft}
%
    \begin{subfigure}[b]{\linewidth}
       \includegraphics[width=\textwidth]{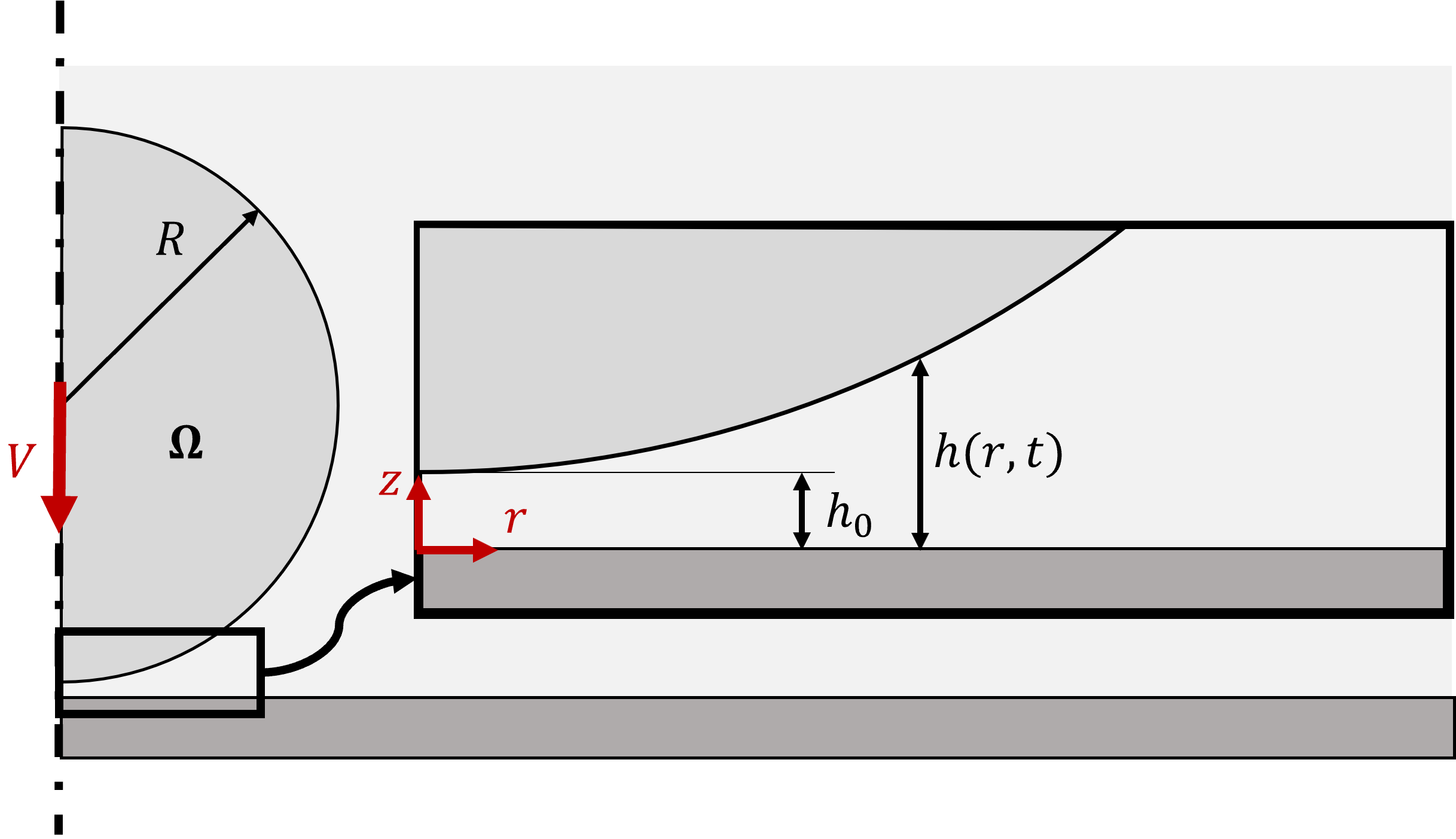}
       \caption{\large  (a)} \label{fig:impact_configuration}
    \end{subfigure}

    \begin{subfigure}[b]{\linewidth}
       \includegraphics[width=\textwidth]{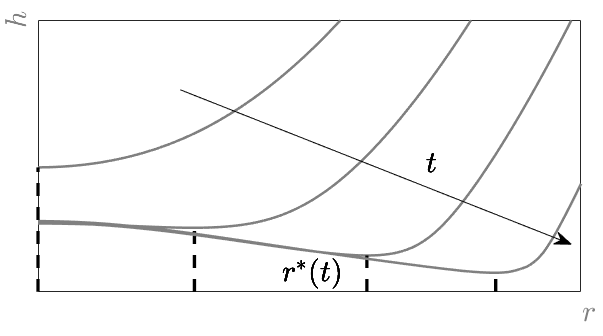}
       \caption{\large  (b)}  \label{fig:radius_approach}
    \end{subfigure}
   
        \caption{(a) Schematic of the impact as modeled at the beginning of the simulation. 
    The coordinate system is set such that the top flat surface of the rigid body (dark gray) is at $z=0$. The solid ball (light gray) approaches the surface with initial velocity $V$ starting with an initial gap $h_0$.
    The surrounding fluid cushions the impact.
    (b) Schematic of the free surface deformation ($\partial \Omega^{-}$) of the impactor while approaching the surface.
    At each time step the point closer to touchdown is identified and its radial coordinate is measured  ($r^*(t)$). The space between this lowest point and the surface is sometimes said to form a ``\textit{neck}'' \cite{Gordillo:2022} or a ``\textit{kink}'' \cite{Ruiter:2012}.}
    \label{fig:schematics}
\end{figure}

The role played by the mediating air is far less clear for soft elastomer impact, and the stress distribution during the impact process remains unknown.
The literature contains examples of a variety of configurations involving solids, 
ranging from a rigid sphere denting an elastic substrate mediated by air \cite{balmforth:2010}, to two spheres immersed in water contacting at relatively low velocity \cite{David:1986}. 
Recently, experiments featured soft solid impactors hitting a microscopically-flat surface \cite{Zheng:2021}.
The maximum \textit{entrapped air bubble} was measured at the first instances of contact, and it was shown that its radius grows as a power-law function of the impact velocity up to a maximum value and, then, drops suddenly and becomes weakly dependent on impact velocity.
It was then suggested that there exist two domains, dominated respectively by elastic and inertial response. 
In both cases, there is a local viscous pressure ``\textit{build up}'' as the air is squeezed, which leads to the formation of a ``\textit{dimple}'' around the tip.
As time progresses the dimple expands and we define the radial position of its edge the ``\textit{deformation front}'' (see \Cref{fig:schematics}). 

Here we provide a scaling analysis across a wide range of impact velocities that include the observed sharp transition between elasticity-dominated and inertia-dominated impacts \cite{Zheng:2021}.
The scaling analyses complements and provides context for numerical simulations of the impact dynamics, which provide additional insight into the pressure in the mediating air film: a pronounced spike in the air pressure emerges as the impact transitions to the inertia-dominant regime. 
The growth or decay of this pressure spike governs whether the solid will glide over the surface, or get sufficiently close to the surface that contact formation cuts off the gliding process, ultimately reducing the lateral scale of the entrained lubricant. 
{\color{black}
Unlike some of the literature in impacts of drops 
\cite{Mandre:2009,Duchemin-Josserand:2011,Gordillo:2022,Sprittles:2023}, in this contribution we do not intend to quantify the local thickness of the thin film, we focus on understanding the evolution of the radius of the entrapped bubble.}
Using a numerical framework, we explore the role of solid compressibility by varying the Poisson's ratio of the solid phase. We find that compressibility in the solid phase can reduce the intensity of the pressure spike and thus modulate both the magnitude and transitional velocity between the elasticity-dominated and inertially-dominated regimes. 

The text is organized as follows: 
Section \ref{sec:scaling} analyzes the relative orders of magnitude of the phenomena involved in the process, 
and Section \ref{sec:methods} briefly introduces our numerical model. 
Section \ref{sec:validation} compares the numerical model's predictions with experiments, scaling analysis, and previous literature. 
Finally, Section \ref{sec:discussion} explains why there is a maximum bubble radius and why its size can change suddenly with model parameters. It also highlights similarities and differences with droplet impacts.
We close with final remarks, including research outlook, in Section \ref{sec:final}.

\section{Scaling}
\label{sec:scaling}

\subsection{Impact setting}
\label{sec:scaling_intro}

A solid sphere of radius $R$ approaches with velocity $V$ a rigid flat surface with normal $(0,0,1)$ in the axisymmetric coordinate system $(r, \theta, z)$.
The surface is located at $z=0$ and its distance to the impactor (gap) can be expressed as:
\begin{equation}
    h(r,t) = z_i(t) 
    + 
    {r^2 \over 2 R} 
    + 
    w(r,t) 
    + 
    \mathcal{O}\left[\left({r\over R}\right)^4\right] \, ,    
\label{eq:parabolic_approx}
\end{equation}
where the first term corresponds to the central line solid position of the impactor in the absence of any solid deformation and $w(r, t)$ is the vertical deformation of the impactor surface.

As the sphere is surrounded by a fluid, when $h$ is sufficiently small, the flow between the leading edge and the flat surface can be modeled with Reynolds lubrication equations \cite{szeri:2010}: 
\begin{align}
         r {\partial (\rho_m h) \over \partial t} 
    =
    {\partial \over \partial r} 
    \left(
    {r \rho_m h^3 \over \mu_m'} 
    {\partial p_m
    \over 
    \partial r} 
    \right) \, ,
    \label{eq:reynolds}
\end{align}
where $\mu_m' = 12 \mu_m$, $\mu_m$ is the dynamic viscosity of the mediating fluid \cite{savitski:2002},
and $\rho_m$ is the density.
The lubrication pressure is directly related to the stress in the solid impactor, $\vecsym{\sigma}$, via the boundary condition of force equilibrium at the interface (normal vector $\vecsym{n}$)
\begin{align}
    (\vecsym{\sigma} \vecsym{n}) \cdot \vecsym{n} 
    = 
    p_m \, .
    \label{eq:bc}
\end{align}

If this pressure is high enough, it can locally halt the approach of the sphere and contact will be made on an annulus, with a dimple of air trapped in the middle. 
The condition for this air pocket formation provides the following relation between the characteristic radius of the dimple $\mathcal{L}$, the one of the impactor $R$ and the characteristic gap-size $H$ (see Appendix \ref{sec:app_dimensionless_groups}): $\mathcal{L}=\sqrt{R H}$ \cite{langley:2019}.

\subsection{Dominant inertia in the impactor}
\label{sec:scaling_inertial}

Enforcing incompressibility, \cref{eq:reynolds} leads to lubrication pressure $P_{\text{lub}} \sim \mu_m' V R /H^2$ \cite{szeri:2010}. 
Provided elasticity is negligible, the inertia of the solid implies $P_{\text{in}} \sim \rho_i V \mathcal{L} / \tau$, with $\rho_i$ the solid density. 
Balancing $P_{\text{in}}$ with $P_{\text{lub}}$ and using the geometrical relation between $\mathcal{L}$ and $H$, one obtains: 
\begin{align}
    &\mathcal{L}_{\text{in}} = 
    R \, \mathrm{Stk}^{-1/3}, 
    &H_{\text{in}} = R \, \mathrm{Stk}^{-2/3} \, ,
    \label{eq:inertial_scales}
\end{align}
with the subscript ``in'' referring to inertia regime. Thus, we recover the parameters in \cite{smith:2003}, where $\mathrm{Stk} = \rho_i V R /\mu_m'$ is the particle Stokes number. 

\subsection{Dominant elasticity in the impactor}
\label{sec:scaling_elastic}

Conversely, if elasticity is dominant (subscript ``el'' hereafter), the characteristic value of strain is $H/\mathcal{L}$, and the pressure scale is thus $P_{\text{el}} \sim G H/\mathcal{L}$, where $G$ is the shear modulus of the impactor. 
Balancing this pressure with $P_{\text{lub}}$ yields
\begin{align}
    &\mathcal{L}_{\text{el}} 
    = 
    (R^4 \mu_m' V/G)^{1/5}, 
    &H_{\text{el}} = (R^{3/2} \mu_m' V/G)^{2/5} \, ,
    \label{eq:elastic_scales}
\end{align}
where we recover the same power laws as \cite{David:1986} and 
\cite{Zheng:2021}, the difference with the latter being that we used the shear modulus instead of the equivalent Young's modulus for ease of interpretation of later results, which amounts to removing an order-one prefactor.

\subsection{Transition between elasticity-dominated and inertia-dominated impactor response}
\label{sec:scaling_transition}

The crossover between elastic and inertial regimes occurs when $P_{\text{in}} \sim P_{\text{el}}$. 

Considering the characteristic value of elastic pressures in the inertial regime by substituting \cref{eq:inertial_scales} in the expression for $P_{\text{el}}$, the transition occurs when
$\phi_{\text{in} \to \text{el}} =  P_{\text{in}}/P_{\text{el}} = (R^2 \rho_i^5 V^8/ G^3 \mu_m'^2)^{1/3} \sim 1$. 

Analogously, substituting \cref{eq:elastic_scales} into $P_{\text{in}}$ we find the characteristic value of inertial forces in the elasticity-dominated regime. 
The transition parameter in this case is 
$\phi_{\text{el} \to \text{in}} =  P_{\text{in}}/P_{\text{el}} = (R^2 \rho_i^5 V^8/ G^3 \mu_m'^2)^{1/5} \sim 1$.

We notice that:
\begin{align}
    \phi_{\text{in} \to \text{el}}^{1/2} =
    \phi_{\text{el} \to \text{in}}^{5/6} = 
    \phi &= 
    {V \mathrm{Stk}^{1/3} \over c_s} \nonumber\\
    &=
    {\mathcal{L}_{\text{in}} / c_s \over H_{\text{in}} / V}
    = {\tau_{\text{propagation}} \over \tau_{\text{impact}}}  \,
    \label{eq:transition_parameter_in_el}
\end{align}
where $c_s= \sqrt{G/\rho_i}$ is the shear wave speed in the solid. Recall that, for most materials, the shear and Rayleigh surface wave have comparable magnitude, $c_s \approx c_R$. 
The transition parameter $\phi$ can be interpreted as the ratio of the two characteristic time scales of the phenomenon: 
$\tau_{\text{impact}} = H_{\text{in}} / V$ represents the impact characteristic timespan during the viscous response phase 
and $\tau_{\text{propagation}} = \mathcal{L}_{\text{in}} / c_s$ is the characteristic time it takes a wave emanating from the tip of the impactor to reach the extent of the region where the viscous pressures develop \cite{hicks+purvis:2013}, i.e., where the deformation accumulates and the dimple forms.

%%%%%%%%%%%%%%%%%%%%%%%%%%%%%%%%%%%%%%%%%%%%%%%%%%%%%%%%%%%%%%%%%%%%%%%%%%%%%%%%%%%%%%%%%%%%%%%

\subsection{Role of solid compressibility}
\label{sec:scaling_solid_compressibility}

The experiments in \cite{Zheng:2021} featured almost-incompressible solid impactors. 
The P-wave modulus of a solid with some compressibility, no matter how little, is $M= 2 G (1-\nu)/(1-2 \nu)$  (here $\nu$ is the Poisson's ratio of the material)
In an uniaxial strain scenario, it is related to the pressure by $P_M \sim M \varepsilon_{zz}$.
The vertical displacement is order $\sim H$, while the part of the impactor which undergoes deformation is within a radius $\sim c_p \tau$ from the tip, where $c_p = \sqrt{M/\rho_i}$ is the P-wave speed within the solid. 
Therefore, we can estimate $P_M \sim \rho_i c_p V$. 
Assuming this pressure balances $P_{\text{lub}}$, we obtain the following scales associated to impactor deformation:
\begin{align}
    &\mathcal{L}_{M} 
    = 
    \left( {R^3 \mu_m' \over \rho_i c_p} \right)^{1/4} \, , 
    &H_{M} 
    = 
    \left( {R \mu_m' \over \rho_i c_p} \right)^{1/2} \, ,
    \label{eq:bulk_scales}
\end{align} 
which do not depend on the impact velocity. 
We can also analyze the conditions for the inertial regime to feature meaningful volumetric deformation by setting $P_M \sim P_{\text{in}}$. 
Pursuing a similar reasoning as the one that led to the definition of $\phi$ (section \ref{sec:scaling_transition}), we evaluate the ratio between $P_M$ and $P_{\text{in}}$ both when inertial scales, \cref{eq:inertial_scales}, are dominant and when those in \cref{eq:bulk_scales} are.
Thus, we obtain a second transition parameter by enforcing them to be of the same magnitude at the said transition:
\begin{align}
    \psi_{in \to M}^{3/4} =
    \psi_{M \to in} = 
    \psi 
    &=
    {V \mathrm{Stk}^{1/3} \over  c_p } \nonumber \\
    &=
    {\mathcal{L}_{\text{in}} / c_p \over  H_{\text{in}} / V}
    = 
    {\tau'_{\text{propagation}} \over \tau_{\text{impact}}}  \, ,
    \label{eq:transition_parameter_compressibility}
\end{align}
which again can be interpreted as the ratio of the time it takes compressive waves to propagate in the solid  ($\tau'_{\text{propagation}}$) over the lubrication time scale in the thin film. 
In addition, $\psi$ is function of $\phi$ as:
\begin{align}
    \psi = {c_s \over c_p} \phi =
    \sqrt{(1 - 2 \nu) \over 2 (1-\nu)} \phi \, .
\end{align} 
The fact that $\psi < \phi$ implies that solid compressibility can only play a role in the inertial regime. 
Moreover, the transition (the point at which $\psi \sim 1$) depends on the Poisson's ratio value: the closer to incompressibility ($\nu = 1/2$), the greater the value of $\phi$ (e.g., higher impact velocities) for which the solid compressibility would become relevant.   

\subsection{Role of fluid compressibility}
\label{sec:scaling_fluid_compressibility}

The effects of fluid compressibility arise when the pressure becomes of the order of the atmospheric one. 
These could modify the transition picture sketched previously. 
The critical height when the compressibility becomes relevant is $H^* = (R \mu_m' V /P_0)^{1/2}$  where $P_0$ is the ambient pressure; unlike \cite{Mandre:2009}, we used $\mu_m'$ in place of $\mu_m$. 
By comparing the dominant pressure scale to the ambient one we obtain so-called ``compressibility parameters'' for each regime:
\begin{subequations}
    \begin{align}
    \epsilon_{\text{el}} 
    &= 
    {P_0 \over P_{\text{el}}}
    =
    {P_0 \over (R^{-1} \mu_m' G^4 V )^{1/5}} \, , 
    \label{eq:compressibilty_parameter_elastic}\\ 
    \epsilon_{\text{in}} 
    &= 
    {P_0 \over P_{\text{in}}}
    =
    {P_0 \over (R \mu_m'^{-1} V^7 \rho_i^4)^{1/3}}
    \label{eq:compressibilty_parameter_inertial}
\end{align}
\end{subequations}
where \cref{eq:compressibilty_parameter_inertial} is the well-known compressibility parameter for high speed droplets \cite{Mandre:2009}, 
and \cref{eq:compressibilty_parameter_elastic} is relevant in the elasticity-dominated scenario.

\section{Methods}
\label{sec:methods}

To validate the theory outlined by the scaling analysis, we use fully-coupled finite element simulations for both the solid and the thin film.

As we expect substantial deformation in the impactor and highly compressive strains, the elastic Saint-Venant model would lead to nonphysical softening \cite{sautter:2022}, therefore we employ the compressible Neo-Hookean hyperelastic model (see Appendix \ref{subsec:simulations}).
This way the computations are more stable, but we stress that results do not vary significantly with respect to the linear-elastic model.
Simulations are performed with the sphere starting at height $h_0 > H_{\text{in}}, \, H_{\text{el}} $. 
Additional details concerning the numerics can be found in Appendix \ref{subsec:simulations}.

We aim to reproduce the experiments reported in \cite{Zheng:2021}, investigating the impact of soft and hard silicone impactors (Zhermack Brand duplication silicones) against a rigid smooth glass surface. 
Material properties are reported in \Cref{tab:mat_impactor} for the impactors, while the mediating air is assigned density $\rho_m = 1.2 \, kg/m^3$ and viscosity $\mu_m = 1.81 \cdot 10^{-5} \, Pa \cdot s$, corresponding to common values at standard pressure and temperature \cite{cimbala:2006}.
\begin{table}[h]
    \caption{Material properties of the two impactors.}
    \label{tab:mat_impactor}
\centering
    \begin{ruledtabular}
        \begin{tabular}{rrr}
         & \textbf{Soft} & \textbf{Hard} \\
         \hline
         \rule{0pt}{3ex}  
        $\rho_i$ [$kg/m^3$] & 1140 & 1140 \\
        $E$ [$kPa$] & 250 & 1100 \\
        $\nu$ & 0.47 & 0.47 \\      
    \end{tabular}
    \end{ruledtabular}
    
\end{table}

Velocities of the impactors span three orders of magnitude, from $0.01 \, m/s$ to approximately $6 \, m/s$. 
Note that $\epsilon_{\text{in}} \sim 1$ around $V \sim 1.5\, m/s$, 
however, we choose to neglect the effects of air compressibility at the expense of losing physical accuracy during the late approach stages.
Including it would require adding advanced stabilization techniques (e.g., streamline upwind/Petrov–Galerkin \cite{Brooks:1982}) in the finite element formulation currently adopted \cite{habchi:2008}, 
yet we shall bear in mind this modeling decision when interpreting the simulations' outcome. 

During the simulations, we track the point on the projectile surface which is closer to contact at any given time and whose radial coordinate we identify as $r^*$. 
We stress that the lowest point may change at each time step.
See \Cref{fig:radius_approach} for a depiction. 
We approximate the initial contact radius with the radial coordinate of the point closest to the rigid surface at the last converged time step $r_{\text{contact}} \approx r^*(t_{\text{final}})$. 
In all cases (see \cref{tab:final_gap}) the local Knudsen number is order 1 at the last converged time step and the error we expect is less than 1\%  (estimated with \Cref{eq:geometrical_law}).

\section{Validation}
\label{sec:validation}

Our first simulations aim at reproducing the experimental results in 
\cite{Zheng:2021} to test the numerical approach; once the latter has been validated, it will be used to verify the main scaling results. Furthermore, we compare our results to equivalent ones in the literature when possible. 

\begin{figure}
    \centering
    \includegraphics[width=.4\textwidth]{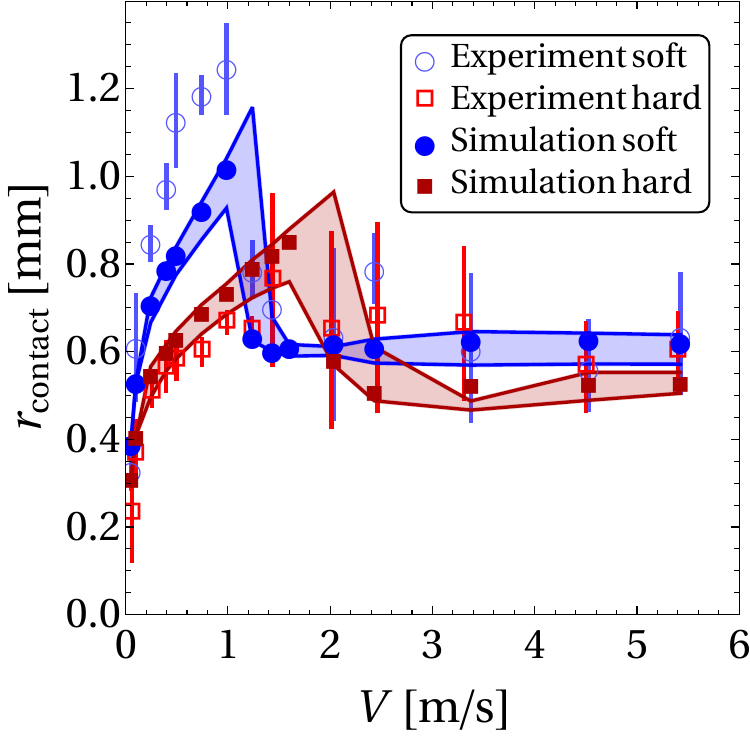}
    \caption{
    The initial contact radius $r_{contact}$ as a function of impact velocity $V$. 
    Experimental data with error bars from \cite{Zheng:2021}
    and numerical data are displayed, for both the soft impactor (circles) and the hard one (squares). 
    The two lines enveloping the shaded area display data from simulations starting from a higher initial gap $2.5 \, h_0$ (bottom lines) and a smaller one $0.5 \, h_0$ (top lines).
    }
    \label{fig:r_sim_exp}
\end{figure}

\Cref{fig:r_sim_exp} shows $r_{contact}$ as a function of $V$ for the experiments performed in \cite{Zheng:2021} and the simulations described in Section \ref{sec:methods}.

In both cases, the radius computed with the numerical model follows a similar trend to the one observed in the experiments. 
The value of $r_{contact}$ increases upon increasing the impact velocity until it reaches a maximum value, and then drops beyond a critical velocity of about $1.1 m/s$. 
For the soft impactor there is a slight underprediction for $V<1.1\, m/s$, while the opposite occurs for the hard impactor.
Remarkably, there is a sudden drop of $r_{contact}$ at values of velocity comparable to those seen in the experiments, even though for the hard impactor the jump is less pronounced.
In both cases, the trapped air bubble radius appears slightly underestimated, albeit within experimental error, when $V>1.1\, m/s$.
The influence of changing the value of $h_0$ is significant only around the sharp transition, where impactors starting closer to the surface entrap more air and manifest the drop at slightly higher velocities.
\begin{figure*}
\centering
\captionsetup[subfigure]{justification=centering}
\begin{subfigure}[b]{.31\linewidth}
   \includegraphics[width=\textwidth]{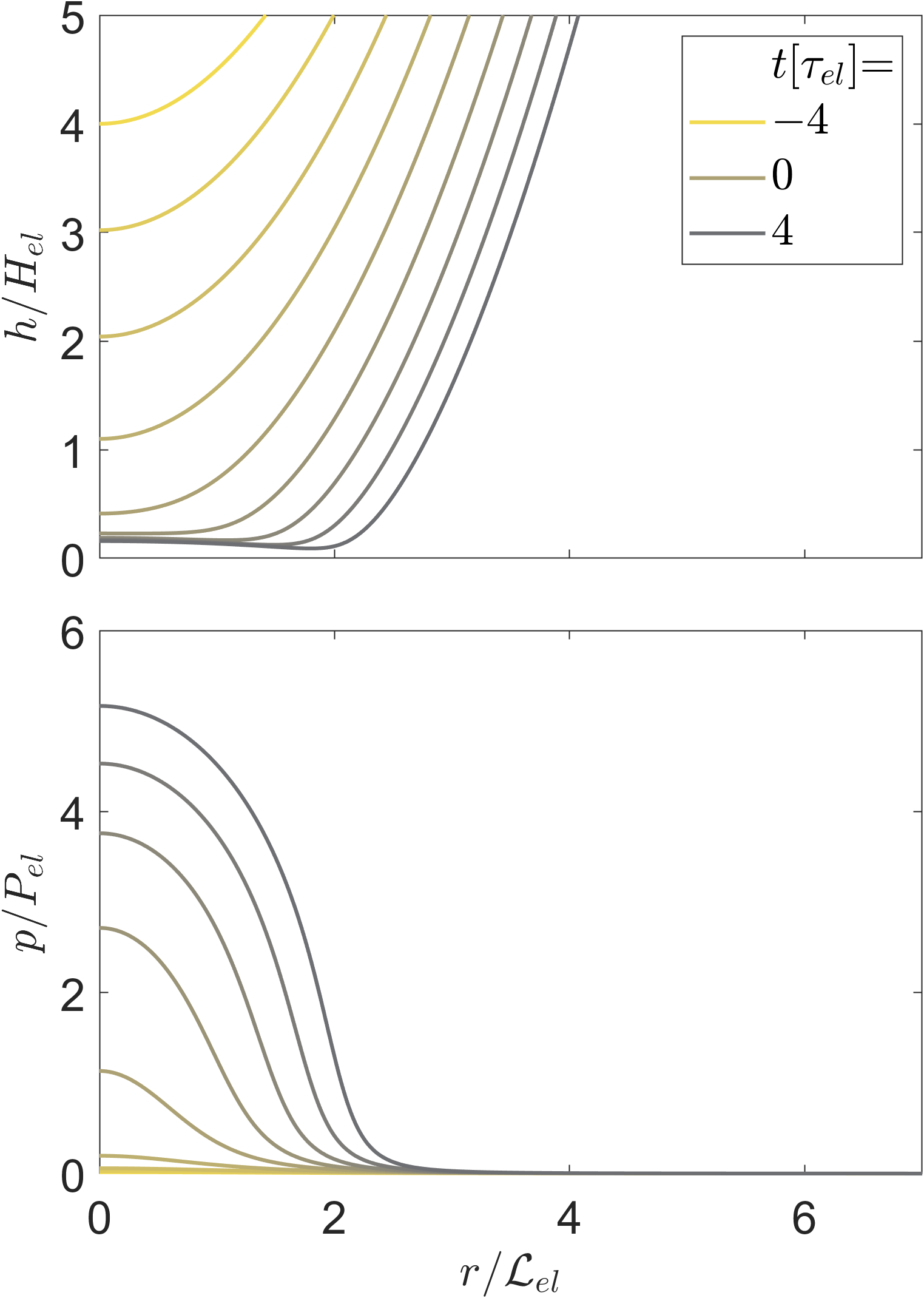}
   \caption{\large  (a) $\phi=0.1$, $\psi=0.02$, $V=0.07 \, m/s$} \label{fig:impact_phi_smaller_than_1}
\end{subfigure}
\begin{subfigure}[b]{.32\linewidth}
   \includegraphics[width=\textwidth]{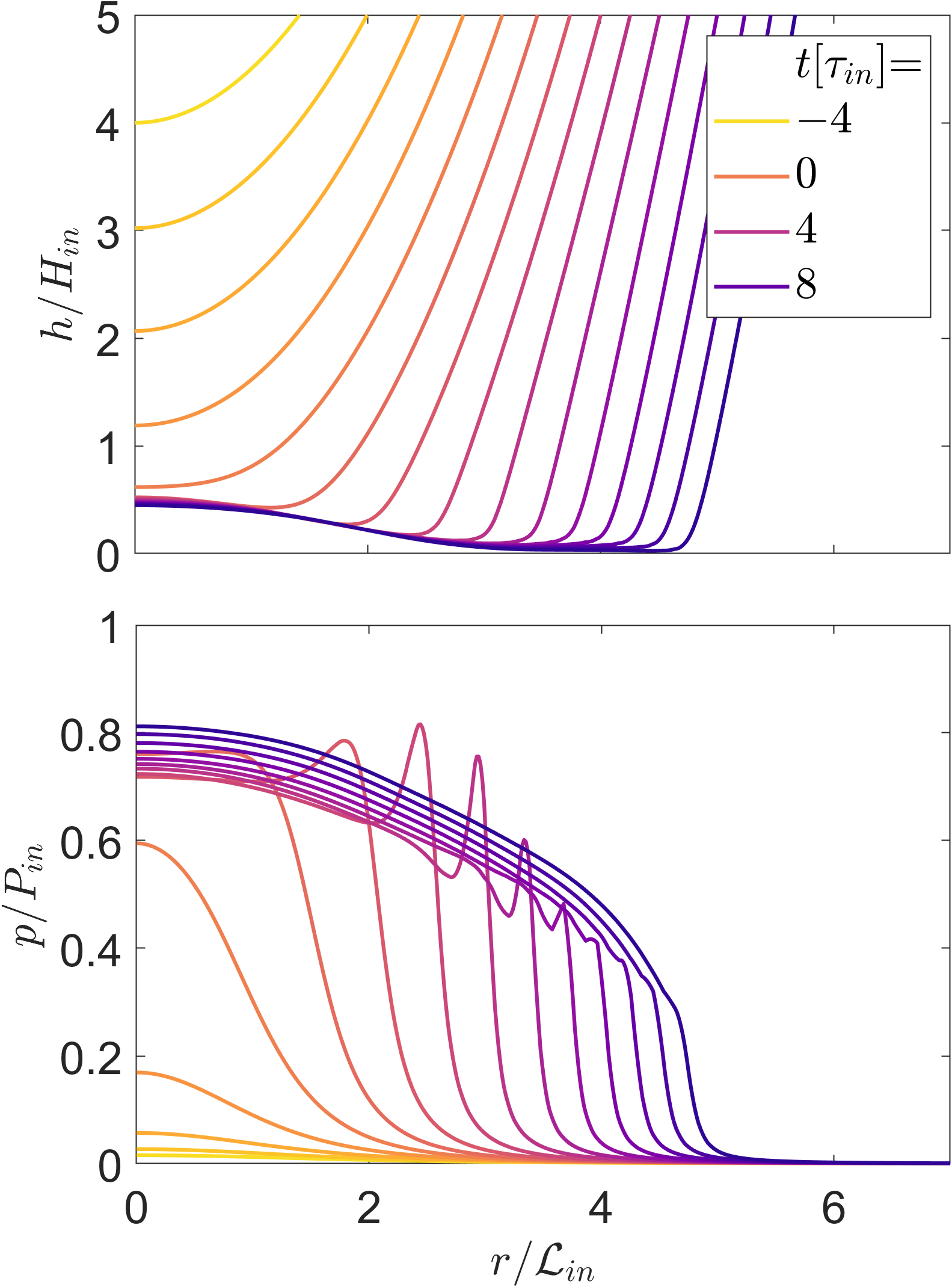}
   \caption{\large  (b) $\phi=4.0$, $\psi=0.95$, $V=1.01 \, m/s$} \label{fig:impact_phi_order_1}
\end{subfigure}
\begin{subfigure}[b]{.32\linewidth}
   \includegraphics[width=\textwidth]{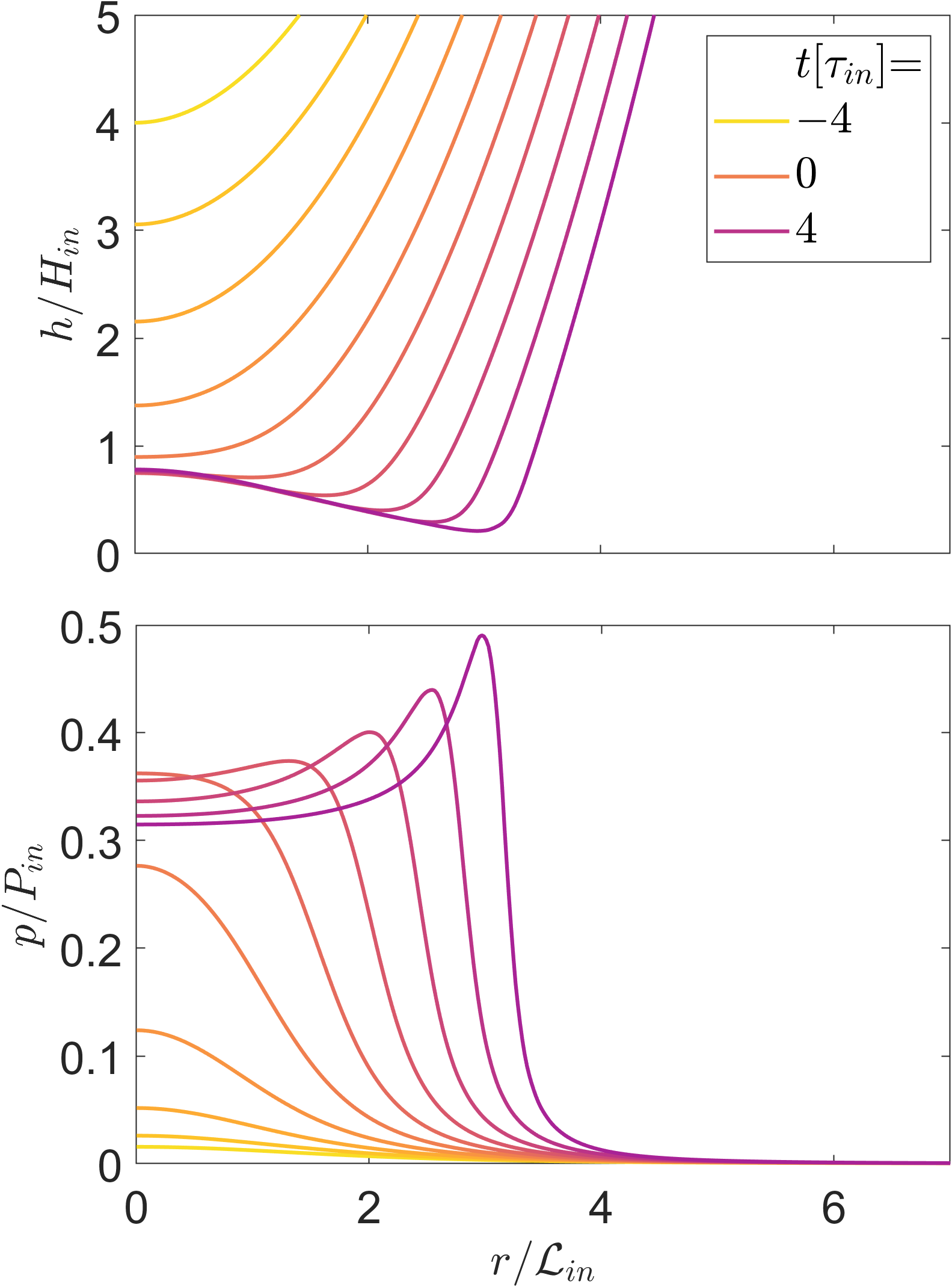}
   \caption{\large (c)
   $\phi = 10.0$, $\psi = 2.4$, $V=2.05 \, m/s$} \label{fig:impact_phi_greater_1}
\end{subfigure}
 \caption{Evolution of the impactor free surface (top) and pressure profile (bottom) during early stages of the impact process. 
 Time is made dimensionless by using the characteristic times $\tau_{\text{in}} = H_{\text{in}}/V$ and $\tau_{\text{el}} = H_{\text{el}}/V$ for the inertial and elastic regime respectively.
 Profiles computed for the soft impactor at low (a), intermediate (b) and high (c) velocity.
 The elastic regime characteristic values are used to make the data dimensionless when $\phi<1$, the inertial regime ones otherwise (see \Cref{tab:physical_values_regimes}). 
 Profiles at $t=0$ would correspond to the time of touchdown if fluid cushioning was absent.
 Low values of $\psi$ reflect volumetric compressibility to be small ($\nu = 0.47$ in all these simulations).}
\label{fig:impact_regimes_pressure_profile}
\end{figure*}
\begin{table}[h]
    \centering
     \caption{Physical values of the characteristic scales for the three impact scenarios presented in \Cref{fig:impact_regimes_pressure_profile}.
     By coincidence the lateral and vertical scales for configuration (a) and (b) have similar values.}
    \label{tab:physical_values_regimes}
    \begin{ruledtabular}
    \begin{tabular}{ccccc}
        & Config. (a) & & Config. (b) & Config. (c)\\ \hline
       $\mathcal{L}_{\text{el}} \, [mm]$  & $0.21$ & $\mathcal{L}_{\text{in}} \, [mm]$ & $0.21$ & $0.17$ \\       
       $H_{\text{el}} \, [\mu m]$  & $6.3$ & $H_{\text{in}} \, [\mu m]$& $6.3$& $3.9$\\
       $P_{\text{el}} \, [KPa]$  & $2.5$ & $P_{\text{in}} \, [KPa]$ &  $39 $ & $203 $\\
       $\tau_{\text{el}} \, [\mu s]$  & $97$ & $\tau_{\text{in}} \, [\mu s]$ & $6.3$ & $1.9$\\
    \end{tabular}    
    \end{ruledtabular}
    
\end{table}

Free surface and pressure profiles during the approach are shown in \Cref{fig:impact_regimes_pressure_profile} for values of the impact parameter corresponding to the elastic regime $\phi=0.1$, the transition $\phi=4$, and the inertial regime $\phi=10$.
In order to verify the corresponding scalings, the data is made dimensionless with the values presented in \Cref{tab:physical_values_regimes}.
It is readily acknowledged that the geometrical length scales of the dimple, the time scales, and pressure scales are of the order of magnitude predicted by the scaling in all three scenarios \footnote{We used the inertial regime scales to adimensionalize the transition, but using the elastic regime's ones would yield similar results}.

\begin{figure*}
\centering
\captionsetup[subfigure]{justification=centering}
\begin{subfigure}[b]{.27\linewidth}
   \includegraphics[width=\textwidth]{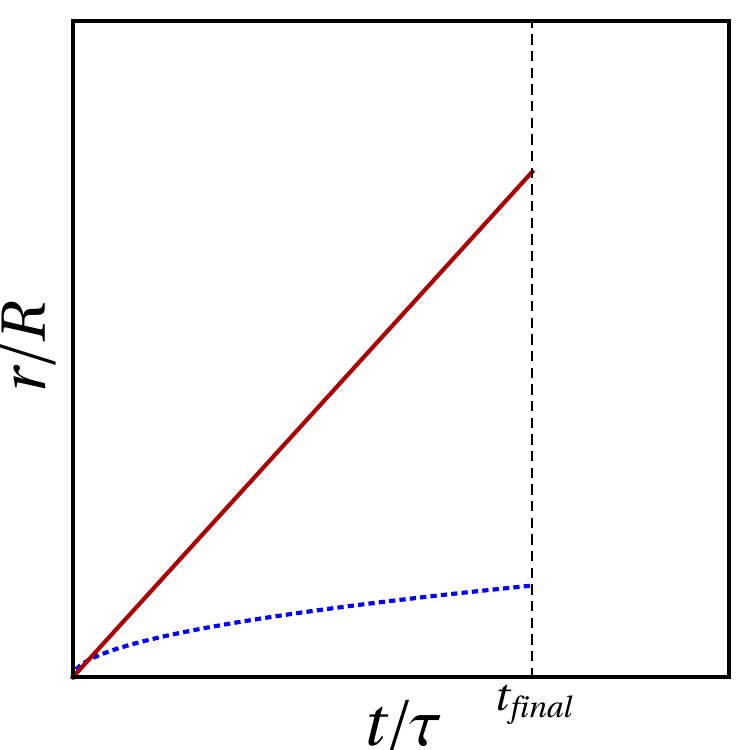}
   \caption{\large  (a) $\phi \ll 1$} \label{fig:scheme_phi_smaller_than_1}
\end{subfigure}
\begin{subfigure}[b]{.27\linewidth}
   \includegraphics[width=\textwidth]{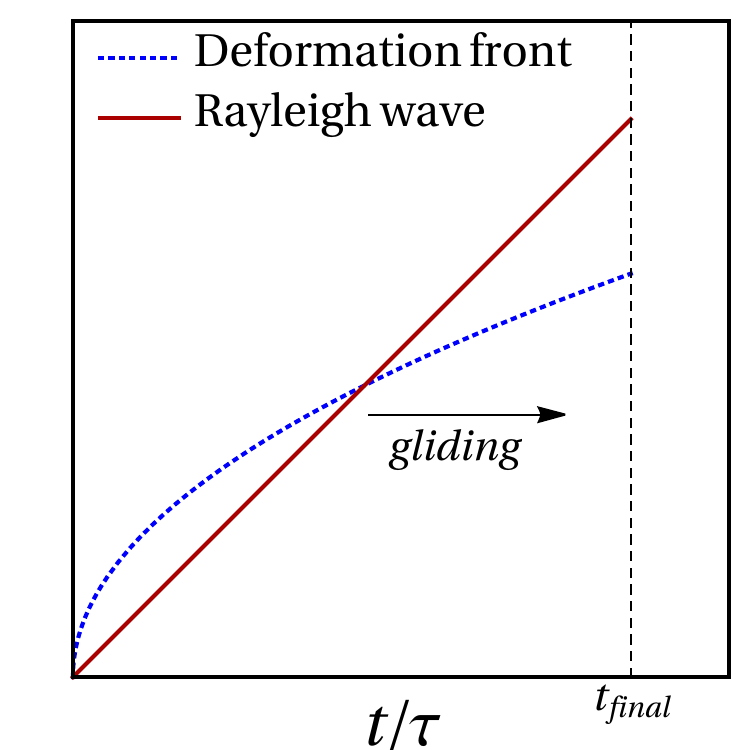}
   \caption{\large  (b) $\phi \sim 1$} \label{fig:scheme_phi_order_1}
\end{subfigure}
\begin{subfigure}[b]{.27\linewidth}
   \includegraphics[width=\textwidth]{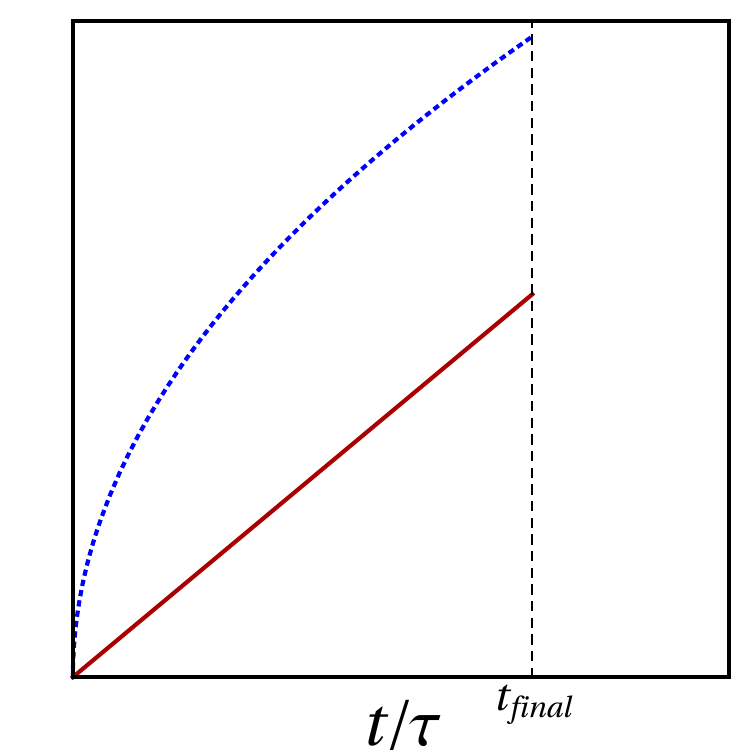}
   \caption{\large (c) $\phi \gg 1$} \label{fig:scheme_phi_greater_1}
\end{subfigure}
 \caption{Schematic comparison over time of radial position of the surface waves v. edge of deformation front, for the three potential behaviors. (a) $\phi \ll 1 \to$ waves propagating much faster than deformation front: slow, elasticity-dominated response; (b) $\phi \sim 1 \to$ waves catch up with deformation front: transition regime, elastic and inertial forces of similar magnitude, gliding; (c) $\phi \gg 1 \to$ waves propagating much slower than deformation front: fast, inertia-dominated response.
 }
 %Great Good!
\label{fig:impact_regimes_cartoon}
\end{figure*}

In the elastic regime (\Cref{fig:impact_phi_smaller_than_1}) the profile is much flatter compared to the inertial one (\Cref{fig:impact_phi_greater_1}), with smoother curvature close to contact.
As the impactor approaches, the pressure reaches its maximum value in the middle and evolves smoothly with time. 
The projectile is deformed accordingly and the point close to contact is significantly slowed down, resulting in low lubrication pressures at the corresponding radial coordinate (see \Cref{tab:physical_values_regimes} for dimensional data as the axis are normalized).
This is the regime studied in \cite{David:1986}, here presented with scales which do not depend on the initial gap $h_0$.

In the inertial regime (\Cref{fig:impact_phi_greater_1}), initially the pressure peaks at $r=0$, slowing down the material points near the axis. 
Information about this deformation propagates at the surface wave speed of the material, which happens to be below the deformation front propagation rate (see Section \ref{sec:deformation_front} and \Cref{fig:scheme_phi_greater_1}).
The impactor response is thus highly local. 
During later stages, the pressure maximum moves away from the center, and, for large times, it is in correspondence with the free surface minima. 
Qualitatively the profiles resemble those observed in numerical simulations of droplets at high speeds \cite{Mandre:2009, Mandre:2010,Hicks:2010}, which is not surprising as the scaling is similar. 

Finally, for the transition there is no prior literature to compare with, but it displays features of both regimes (\Cref{fig:impact_phi_order_1}). 
At first, the response resembles the one of the inertial regime, 
but later on the elastic one. 
This transition is dictated by the impact being initially inertia dominated, developing intense localized forces (pressure spikes) sufficient to halt the fall locally, giving enough time to the surface wave to reach the outward-moving deformation front. 
This postpones the occurrence of contact, and engages larger portions of the solid, compared to either limit cases. 
In other words, the waves emanating from the tip reach the edge of the pressure front, ``warn'' the material ahead about the incoming solicitation, and bring larger portions of solid to bear the load in a concerted deformation to resist the viscous pressures, as it so happens in the elastic regime.  
The corresponding free surface profile shows the ``frozen'' dimple profile and gliding similarly to what has been observed experimentally \cite{langley:2017} for highly viscous drops. 
It also manifests a small ``lift-off'' angle \cite{kolinski:2014} and the ``double kink'' as observed in \cite{Ruiter:2012}.
The gliding stage would be completely averted if the information about the tip deformation transported by surface waves did not reach the deformation front in time. 

The situation for the aforementioned three scenarios is also sketched in \Cref{fig:scheme_phi_smaller_than_1,fig:scheme_phi_order_1,fig:scheme_phi_greater_1}. 
The radial position of the deformation front goes like $\sim t^{1/2}$ (see, e.g., \cite{langley:2017} as well as the discussion in section \ref{sec:deformation_front}).  
When $\phi \ll 1 \implies \tau_{\text{impact}} \gg \tau_{\text{propagation}}$ (left), the information about the impending impact propagates much faster than the deformation front, so the leading edge responds in unison. 
For $\phi \sim 1 \implies \tau_{\text{impact}} \sim \tau_{\text{propagation}}$ (center), initially the deformation front moves outward faster than the surface waves, but these eventually catch up and engage larger portions of material to resist the lubrication pressures that halt the approach. 
Finally, when $\phi \gg 1 \implies \tau_{\text{impact}} \ll \tau_{\text{propagation}}$ (right), the waves do not reach the deformation front before touchdown, the impactor response is very local. 

\section{Discussion}
\label{sec:discussion}

\subsection{Maximum entrapped bubble}
\label{sec:maximum_bubble}

In the previous paragraph we argued that gliding appears when the physical parameters yield $\phi \sim 1$, but
altering their values ever so slightly (e.g., increasing the impactor initial velocity or decreasing the wave velocity) is enough for the deformation front to elude the surface waves (i.e., the ``\textit{signals}'' in charge of making the rest of the leading edge aware of the changing conditions), 
thus preventing gliding altogether. 
Thus, the leading edge quickly and locally punches through the air film, no gliding takes place, and touchdown occurs without delay. 
That is why there is a maximum in the radius of the entrapped bubble, recall \Cref{fig:r_sim_exp}.  

\begin{figure*}
\centering
\captionsetup[subfigure]{justification=centering}
\begin{subfigure}[b]{.4\linewidth}
   \includegraphics[width = \textwidth]{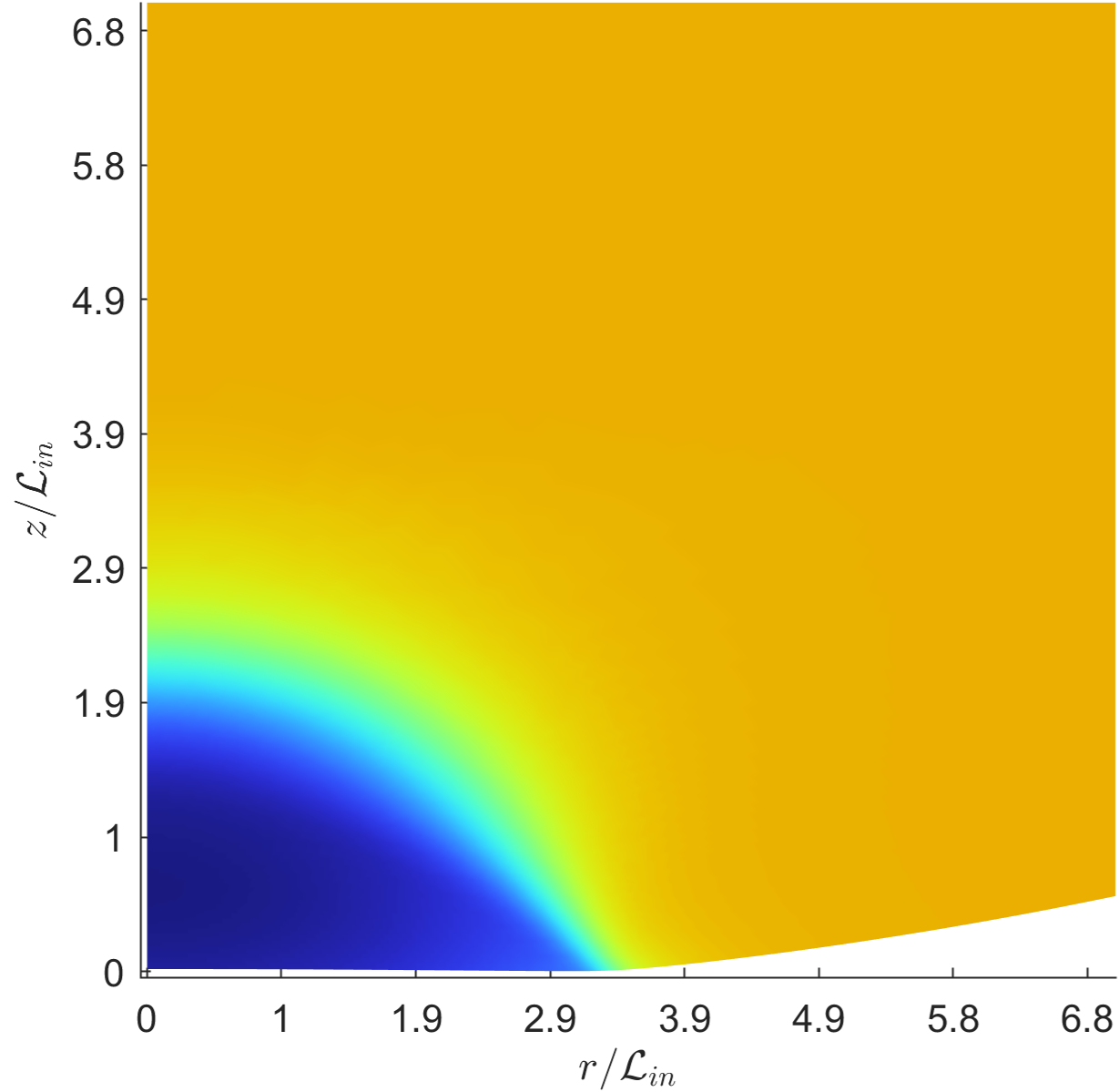}
   \caption{(a) $\nu = 0.1$}
   \label{fig:strain_poisson_01}
\end{subfigure}
\begin{subfigure}[b]{.15\linewidth}
   \includegraphics[width =.45\textwidth]{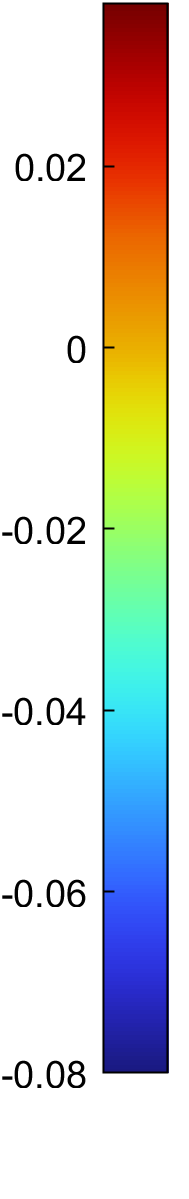}
\end{subfigure}
\begin{subfigure}[b]{.4\linewidth}
   \includegraphics[width = \textwidth]{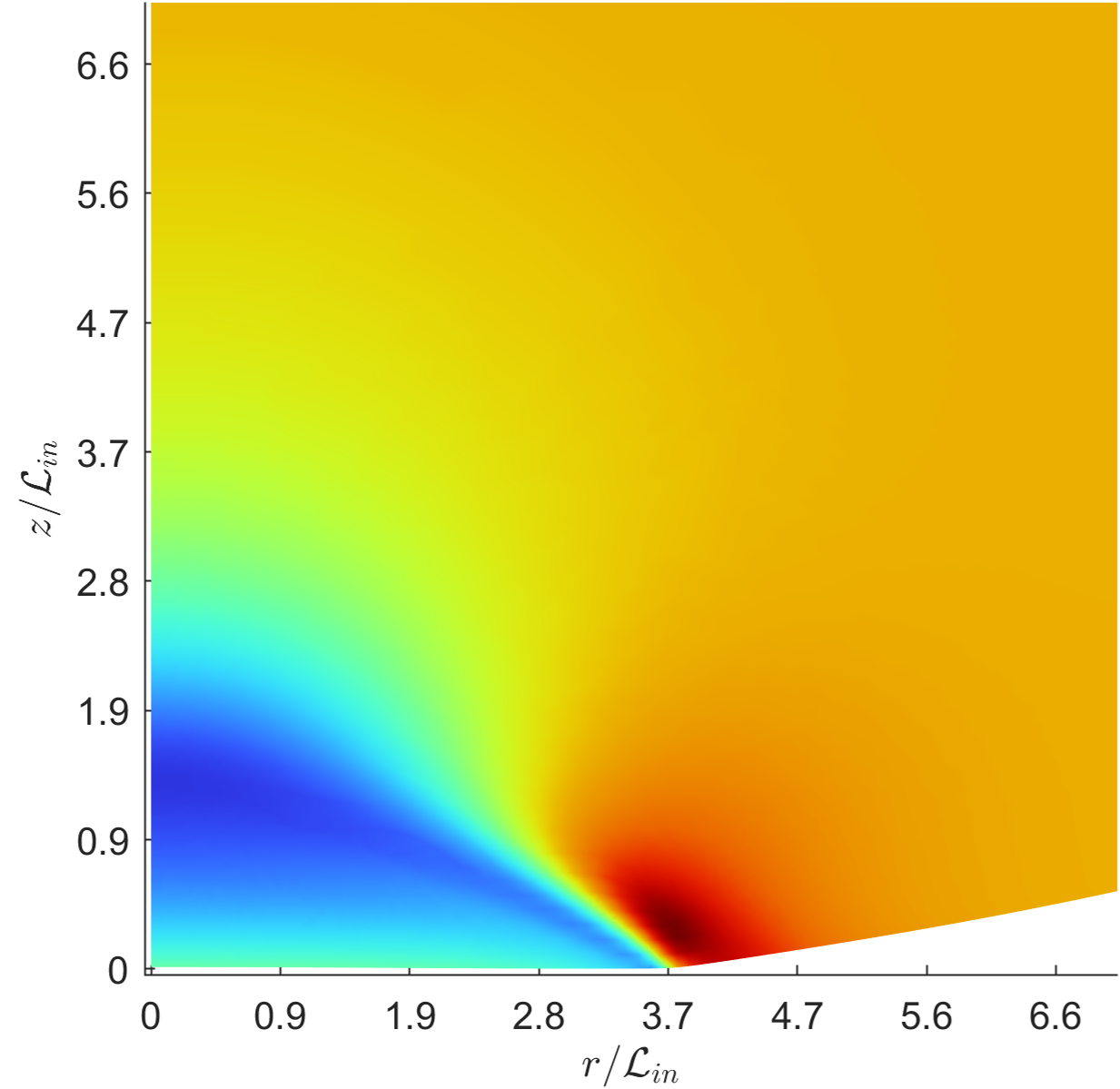}
   \caption{(b) $\nu=0.47$} 
   \label{fig:strain_poisson_047}
\end{subfigure}
\caption{Vertical strain component $\varepsilon_{zz}$ at time $t/\tau_{\text{in}} = 5$ for value of the transition parameter $\phi=3.8$, for a highly compressible solid (left) and a nearly-incompressible one (right).
The compressible case displays higher values of compressive strain, as the material can easily undergo volumetric deformation.
When the bulk modulus is higher, 
as the volume cannot adapt much, the material is partly pushed away by the viscous pressure, causing tensile strain at the edge of the dimple. 
This feature, absent when solid compressibility is relevant, is at the origin of the spikes observed in the pressure profile of the thin film.
}
\label{fig:vertical_strain_comparison}
\end{figure*}

As to the apparent sudden drop at $\phi \sim 1$, we put forth a kinematic explanation in three steps.  
\begin{itemize}
    \item As the tip region is arrested, the rest of the leading edge continues approaching the surface, unaware of the sudden solicitation, until either surface waves or the viscous pressure build-up reach it.  
    \item Just as one would expect for droplets, the nearly-incompressible material pushed by the lubrication pressure not only goes up,  but also affects the bulk around it (\Cref{fig:strain_poisson_047}). 
    Conversely, it could just compress locally if $\nu \approx 0$ (\Cref{fig:strain_poisson_01}). 
    As the outer surrounding material keeps falling while this happens, strain accumulates at the edge of the deformation front. 
    These strains are highly localized and of a greater magnitude than it would otherwise be if local volume changes could be accommodated by the material.   
    This is the origin of the pressure spikes that are seen both in nearly-incompressible solids and in incompressible droplets \cite{Hicks:2010,hicks+purvis:2013}. 
    \item The pressure spikes in turn induce gliding, which delays touchdown. Surface waves catch up with the deformation front 
    extending the process until it finishes in similar fashion to how it does in the elastic regime. 
    Hence, the greater the $\phi$, the faster the approach process unfolds, leaving eventually no time for surface waves to catch up and gliding to occur.   
\end{itemize}
Let us stress that this behavior is thus intimately intertwined with the lack of solid compressibility. 
Further details are provided in Section \ref{sec:volumetric} devoted to effects of allowing volumetric changes in the solid. 
{\color{black}
What we have proposed is a qualitative mechanism to explain the gliding of the solid impactor; however, a detailed description thereof, just like it has been done in the case of droplets \cite{Gordillo:2022}, merits a local analysis of both the deformation and the thin film flow around the region where the gap is minimal (i.e., the ``neck'' \cite{Gordillo:2022}).}

The severity of the sharp transition of the radius depends both on the velocity when $\phi \sim 1$ and on the elastic modulus of the impactor. 
Given two impactors with the same material properties except the stiffness as in Section \ref{sec:methods}, we expect the transition to occur at higher velocity for the harder one, but with a smaller overall reduction of the radius, as confirmed by simulations.

\subsection{Deformation front evolution}
\label{sec:deformation_front}

In order to further substantiate our theory, we study the evolution of the outer edge of the dimple during the impact event.

\begin{figure}[h]
    \centering
    \includegraphics[width=.5\textwidth]{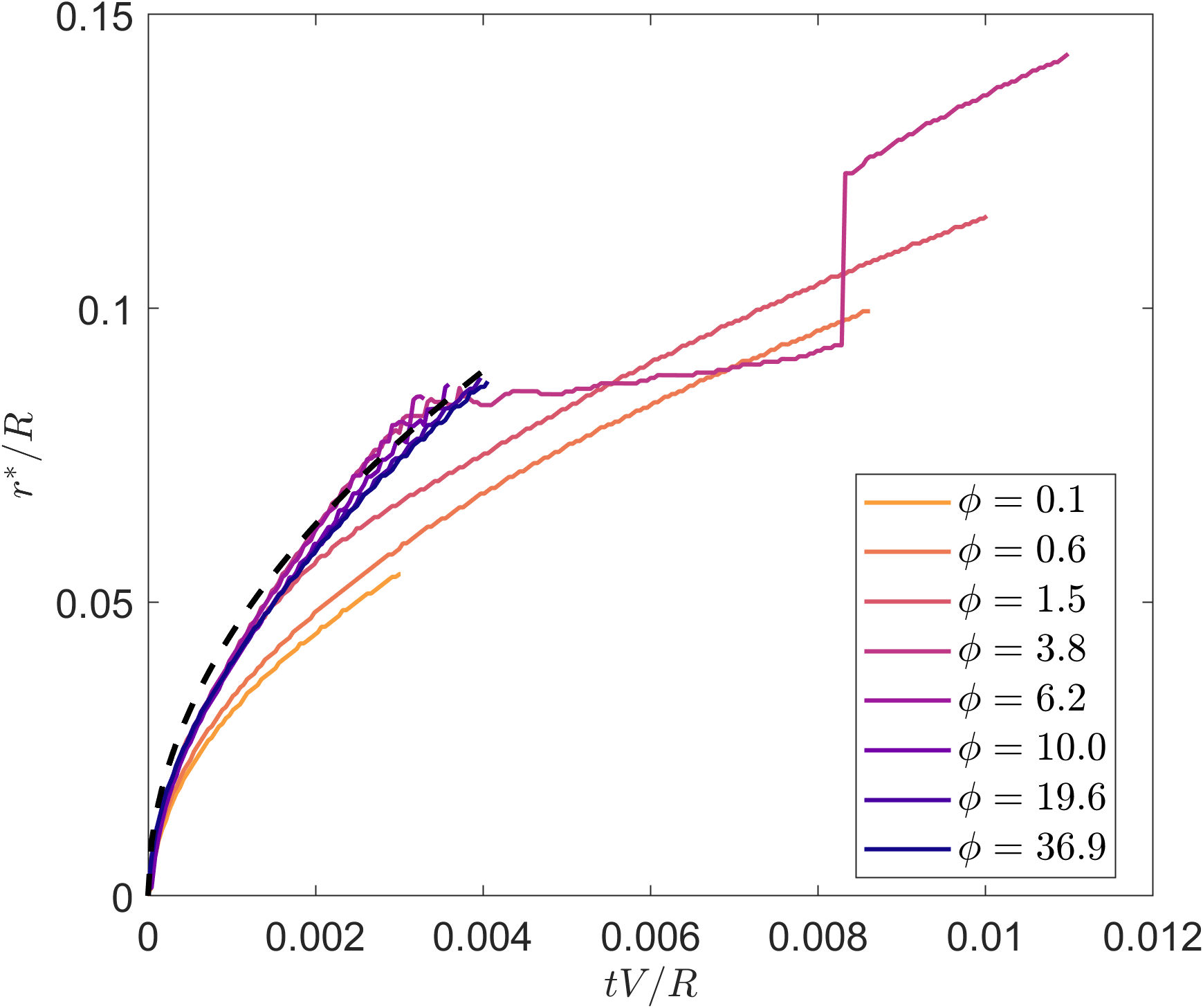}
    \caption{Values of $r^*$ as a function of normalized impact time (cf. Figure 7 in \cite{langley:2017}):  
    numerical results for different values of the transition parameter and square-root-of-time \Cref{eq:geometrical_law} geometrical law (dashed black line). 
    For all curves, the time origin is taken at the instant when $r^*>0$ for the first time.
    Curves in the inertial regime $\phi>4$ collapse onto the geometrical law, whereas as elasticity becomes dominant, so does the gliding of the impactor.
    This plot confirms that at the transition $\phi \approx 4$ the behavior follows initially the inertial one, however gliding occurs abruptly causing a jump in the profile before contact is made, indicating that curvature changes modify the point that is closest to touchdown.}
    \label{fig:radius_time}
\end{figure}

We can use \cref{eq:parabolic_approx} to estimate the lateral velocity of the deformation front. 
Assuming that the viscous response starts at the tip once $h(0,t_0)=H$, leads to
\begin{align}
    h(r,t) = H \approx (H - V \Delta t) + r^2 / 2R   \implies 
    r^* \approx \sqrt{2 R V \Delta t} \, ,     
    \label{eq:geometrical_law}
\end{align}
where $\Delta t$ measures time increment after $t_0$, and we are implicitly assuming by using $r^*$ that this radial position also corresponds to the edge of the dimple.   
In \Cref{fig:radius_time}, simulation data for the soft impactor at different values of the transition parameter are shown.

In the inertial regime, the curves go with $\sim t^{1/2}$ as predicted in \Cref{eq:geometrical_law},
implying that each of the material points is unaware of the deformation of its neighbors.

Since we have a function $r^* = r^*(t)$ that approximates how long it takes for any position of the leading edge to start experiencing the viscous effects from the moment the tip does, we interpret its rate of change as an estimate of the apparent  horizontal ``velocity'' of the front, from the tip to the rest of the interface: 
\begin{align}
\label{eq:velocity_front_estimate}
    \dot{r}^*
    =
    {d r^* \over d t}
    \sim
    \sqrt{R V \over \tau_{\text{impact}}} 
    =
    \mathrm{Stk}^{1/3}\mathrm{V} \, ,
\end{align}   
in agreement with the wetting radius velocity obtained for inviscid droplet impact in a more detailed model in \cite{riboux:2014}. 
\Cref{eq:velocity_front_estimate} means that the deformation front can move much faster than the impact velocity. 
Comparing $\dot{r}^*$ to the wave speed in the impactor offers yet another interpretation of $\phi$, as the ratio between the characteristic velocity of the front over the characteristic velocity of the signals in charge of making the rest of the body aware of the solicitations developing at the tip:
\begin{align}
    { 
    \dot{r}^*
    \over
    c_R
    }
    \sim
    \mathrm{Stk}^{1/3}
    {
    V
    \over
    c_s
    }
    \sim
    \phi \, ,
\end{align}
It follows naturally that for $\phi \ll 1$, the whole body becomes aware of the solicitation and promptly adapts to any lubrication stresses, as the waves propagate over the impactor many times in the span that it takes for the pressure distribution to change significantly. 
Finally, we confirm the assumption that around the transition ($\phi = 4$) the behavior of the impactor initially follows the inertial laws, before collapsing onto the elastic one after extensive gliding.

\subsection{Robustness of the scaling}

\begin{figure}[h]
    \centering
    \includegraphics[width=.5\textwidth]{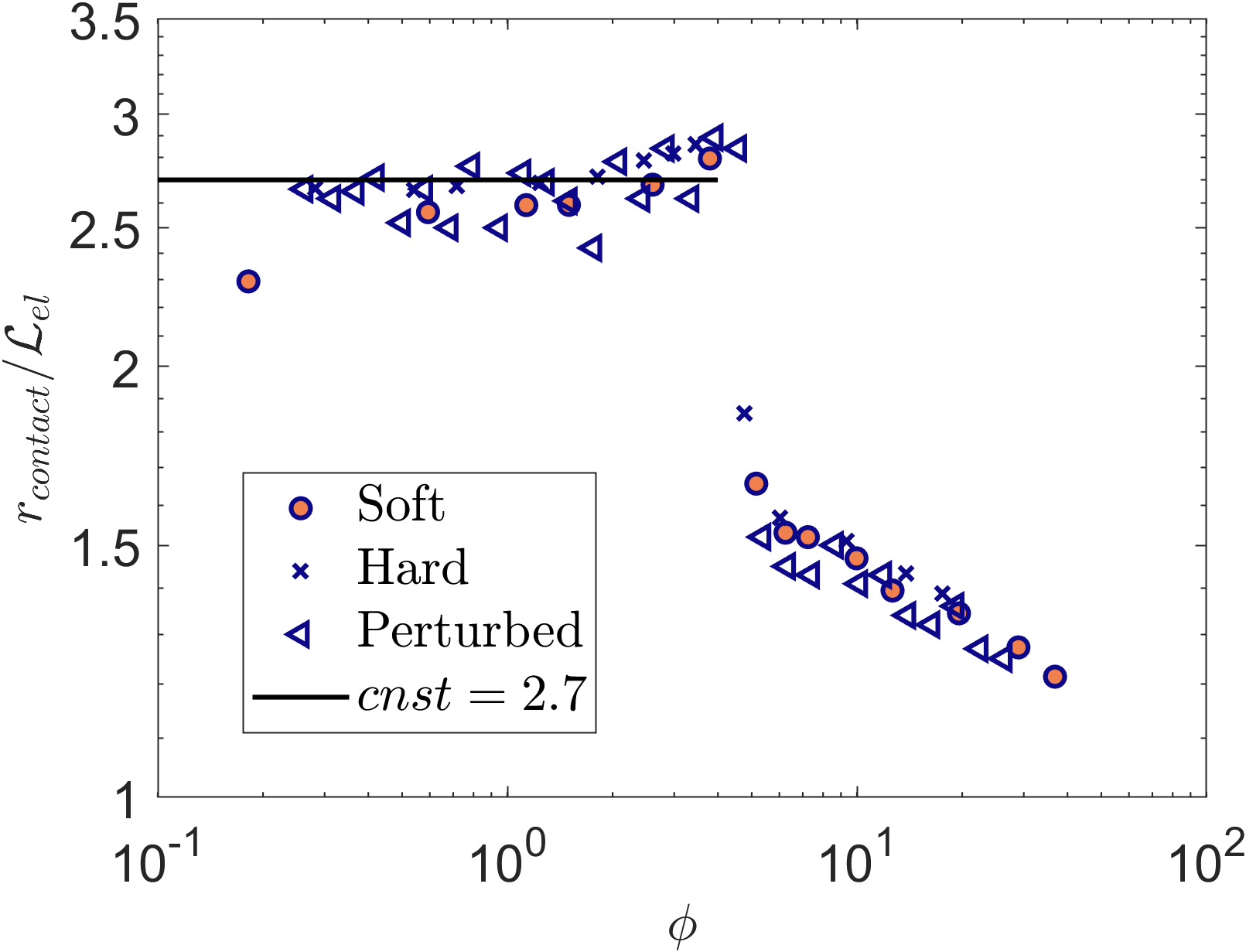}
    \caption{Numerically computed dimple radius expressed in dimensionless form as a function of the transition parameter $\phi$.
    Data for both the ``hard'' and ``soft'' impactor collapses on a master curve.
    Simulations performed with perturbed values of the parameters also show the same trend.
    The sharp drop consistently takes place at values of $\phi \approx 4$. 
    }
    \label{fig:loglog_elastic}
\end{figure}

To check the robustness of the transition parameter we perform additional simulations. 
Taking the values ($\rho_m, \, \rho_i, \, R, \,  \mu_m,$) as specified in Section \ref{sec:methods} and $E= 675 \, kPa$ an intermediate value between the ``soft`` and ``hard`` case, for each value of $\phi$ we vary each of the above parameters independently: given a random number $n$, sampled from the uniform distribution between $(-0.5, 0.5)$, each of the previous parameters is multiplied by $(1+n)$, effectively being altered by $\pm 50 \%$ at most. 
Finally, the impact velocity $V$ is selected to obtain the desired value of $\phi$. 
Notice how in this case we keep the value of the Poisson's ratio constant, as well as $h_0$.
Results are shown in \Cref{fig:loglog_elastic}.

In the elastic regime $r_{contact}$ collapses on the value of $\approx 2.7 \mathcal{L}_{\text{el}}$. 
The sharp transition occurs at $\phi \approx 4$ in all cases. 
For higher values of $\phi$ the data briefly follows the inertial scaling law, before the effect of local elastic compressibility becomes relevant ($\psi \sim 1$). 
The next section will address this issue.

\subsection{Effects of volumetric compression}
\label{sec:volumetric}
\begin{figure}[h]
    \centering
    \includegraphics[width = .45\textwidth]{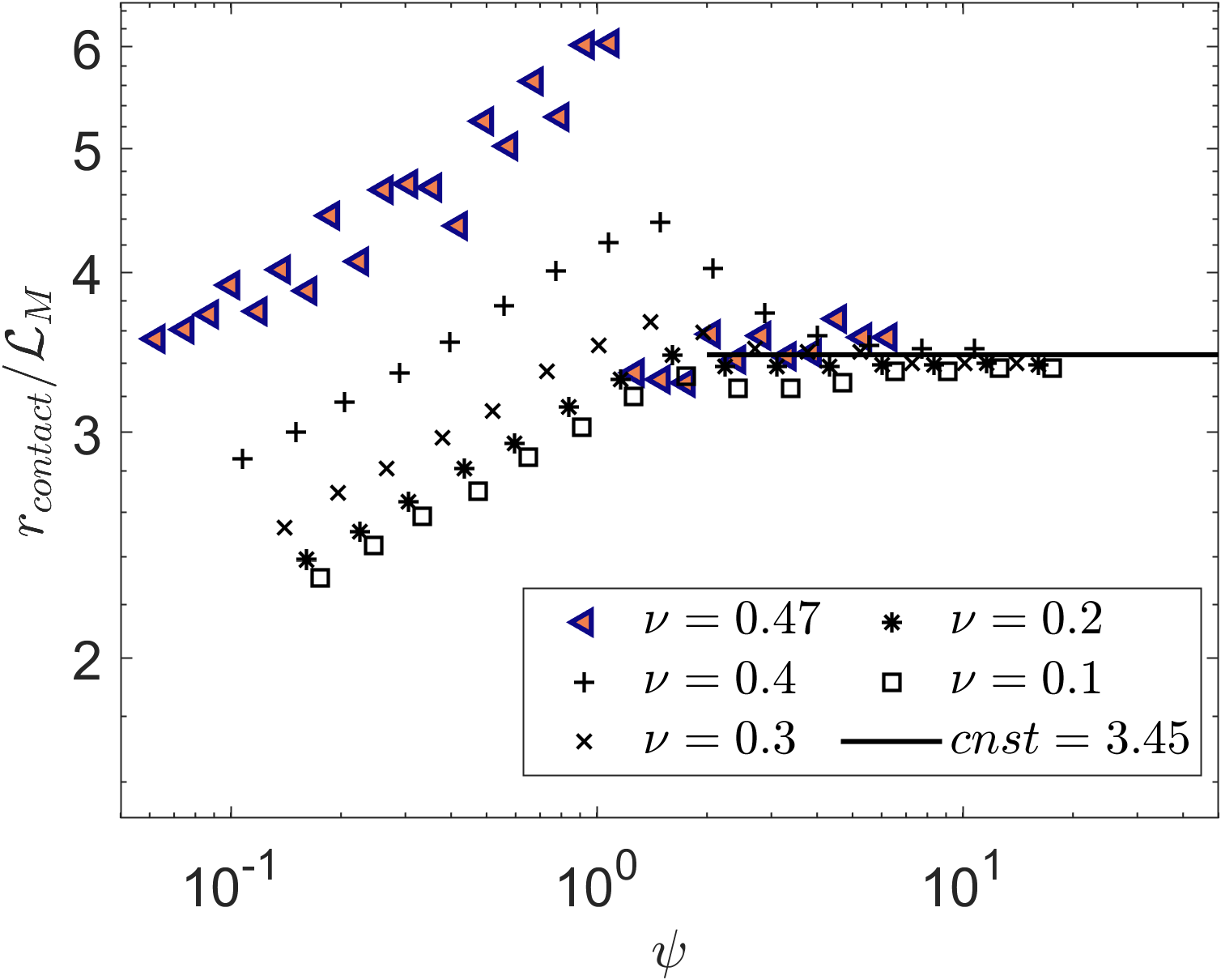}
    \caption{Numerically-computed dimple radius as a function of the transition parameter $\psi$ for different values of Poisson's ratios. 
    For almost-incompressible impactors ($\nu = 0.47$) the sharp drop is evident but it fades away as the solid becomes more easily compressible; by $\nu = 0.2$, the drop disappears. 
    The later scale obtained by balancing the lubrication pressure in the thin film with the volumetric stress in the solid proves correct for values of $\psi>3$.}
    \label{fig:loglog_bulk}
\end{figure}
\begin{figure}[h!]
\centering
\captionsetup[subfigure]{justification=centering}
\begin{subfigure}[b]{.64\linewidth}
   \includegraphics[width=\textwidth]{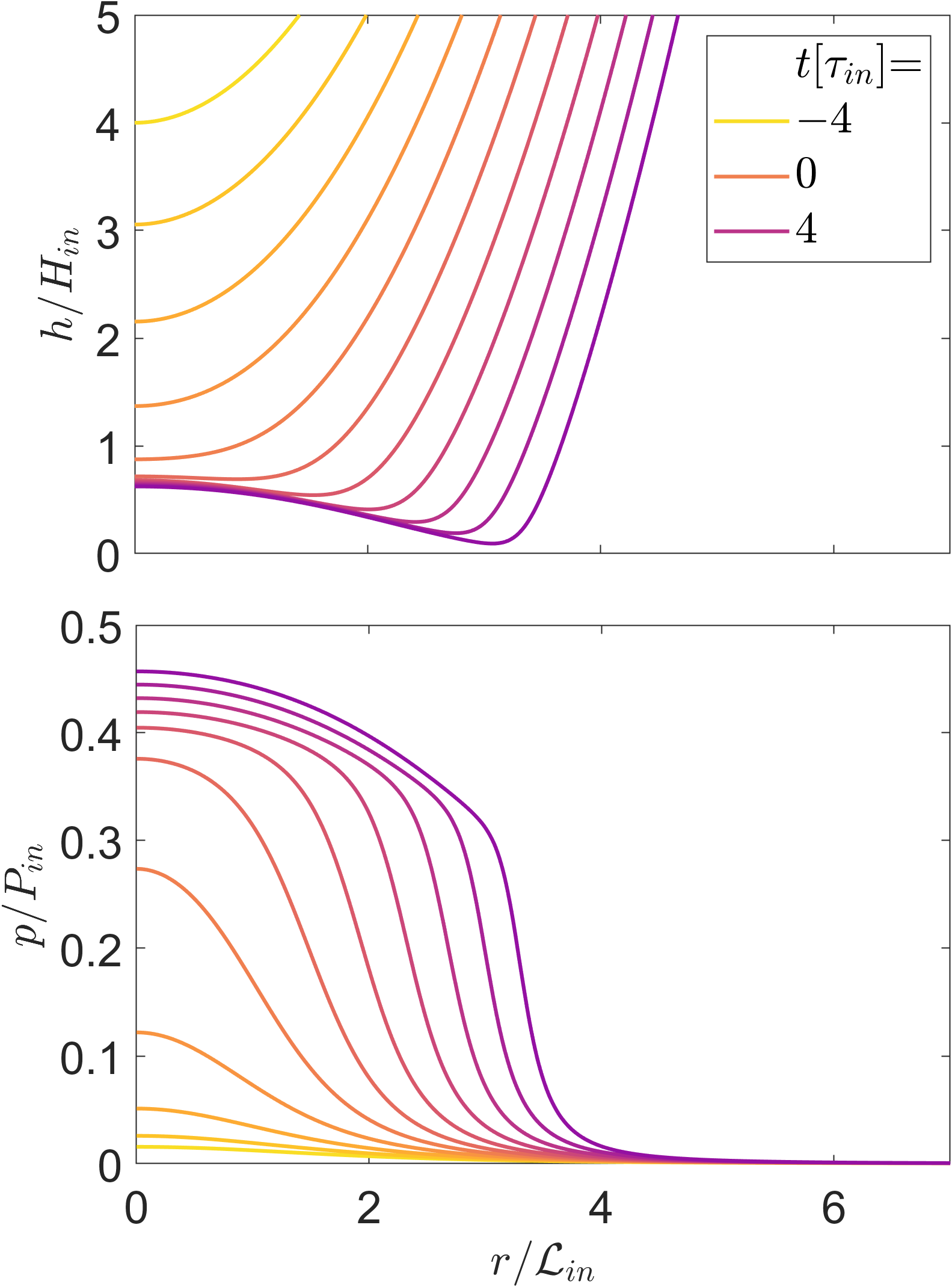}
   \caption{(a) $\phi=4.0$, $\psi=2.7$, $V = 1.1 \, m/s$} \label{fig:impact_M_no_spikes}
\end{subfigure}
\begin{subfigure}[b]{.64\linewidth}
   \includegraphics[width=\textwidth]{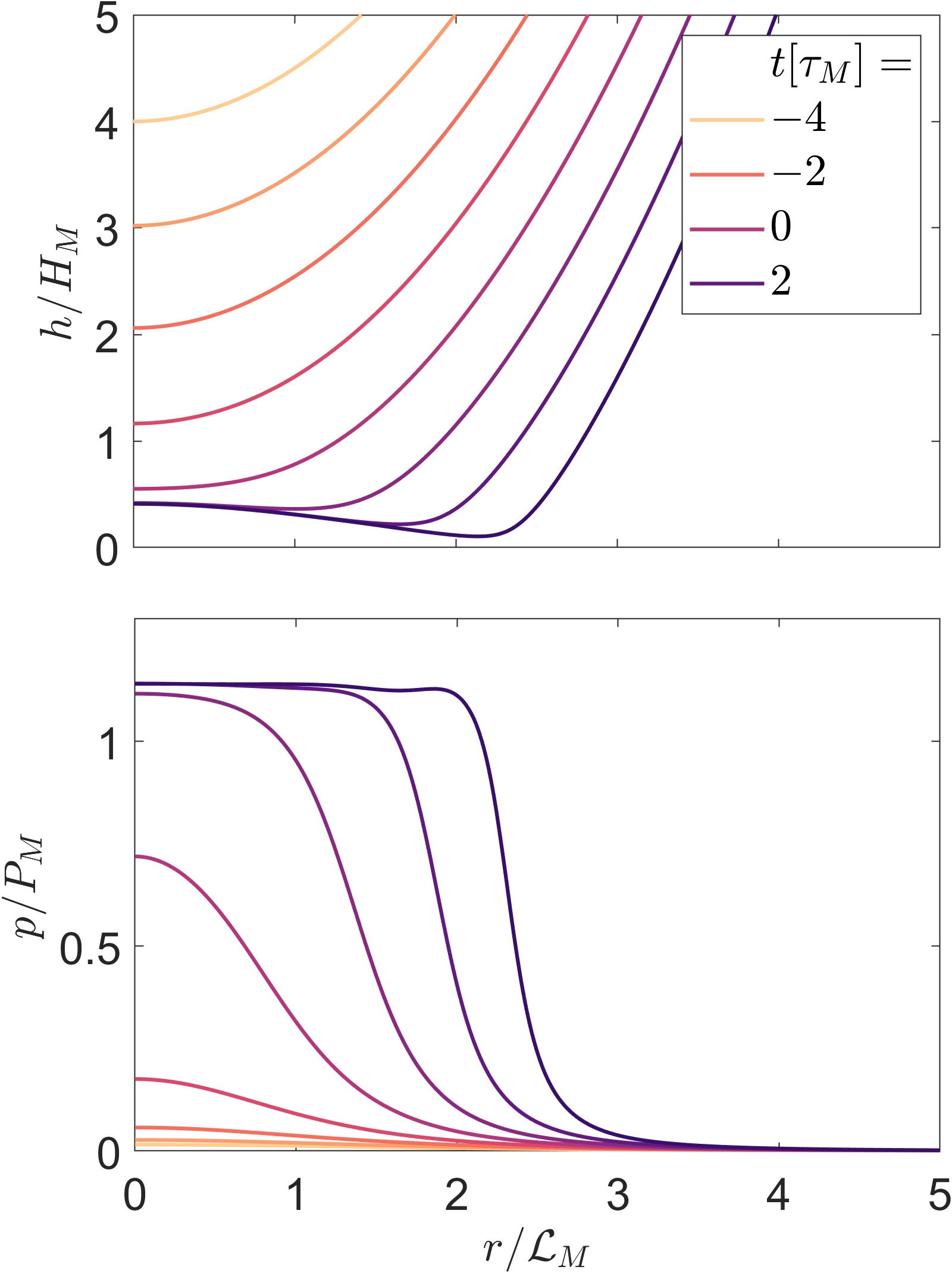}
   \caption{(b) $\phi=30$, $\psi=20$, $V = 5.2 \,m/s$} \label{fig:impact_M}
\end{subfigure}
 \caption{Evolution of the impactor free surface (top) and pressure profile (bottom) during early stages of impact for values of $\nu=0.1$ ($\tau_{\text{in}} = H_{\text{in}}/V$ and $\tau_{M} = H_{M}/V$). 
 On top, the same value of $\phi$ as \Cref{fig:impact_phi_order_1}, to which corresponds a higher value of $\psi$. 
 Compressibility prevents the formation of the pressure spikes which would cause gliding. 
 The time of impact is therefore shorter.
 At the bottom, the profiles in the solid compressible regime. 
 Compared to the inertial case there is flattening of the pressure in the final stages.}
\label{fig:impact_regimes_bulk}
\end{figure}

To better ascertain the mechanism behind the large dimple formation, we perform an additional set of simulations for Poisson's ratio spanning the interval $[0.1-0.4]$, and velocities in $[0.1-10] m/s$, while keeping the other parameters fixed. 
Simulations' outcome of the dimple radius as a function of $\psi$, as defined in \Cref{eq:transition_parameter_compressibility}, can be examined in \Cref{fig:loglog_bulk}.

Remarkably, at values of $\psi > 3$, $r_{contact} \approx 3.45 \, \mathcal{L}_M$ independently of compressibility.
For low values of the Poisson's ratio the bubble radius evolves smoothly and there is no sharp drop in the transition between regimes. 
We have posited that the absence of pressure spikes, which would be immediately released due to volumetric compressibility, prevents the impactor from gliding (\Cref{fig:impact_M_no_spikes}).
Looking at the pressure and free surface profiles for values of $\psi = 20$, where compressibility in the solid is fully dominant (see \Cref{fig:impact_M}),
the free surface profiles are qualitatively similar to those observed for the inertial regime (\Cref{fig:impact_phi_greater_1}), however the pressure profile appear much flatter in the late stages of impact. 
The lengths, time and pressure scales are consistent with the scaling in Section \ref{sec:scaling}. 

Finally, with regard to the timespan as a function of compressibility: 
the case $\nu = 0.1$ lasts until $t/\tau_{in} \approx 5$ (\Cref{fig:impact_M_no_spikes}), 
while $\nu = 0.47$ reaches $t/\tau_{in} \approx 10$ (\Cref{fig:impact_phi_order_1}). 
The time scale $\tau_{\text{in}} = H_{\text{in}}/V$ is the same in both cases, as $H_{\text{in}}$ \cref{eq:inertial_scales} does not depend on $\nu$, which is the only value that changes between these simulations, thus a direct comparison of times in absolute terms is warranted. 
These results hence showcase how near-incompressibility alone can significantly extend the duration of the approach phase, inducing gliding that leads to larger entrapped air bubbles. 

\subsection{Time scale ratio for droplet impact regimes}
\label{sec:droplet_analysis}

We have highlighted throughout this work how the inertial regime for solid impactors is akin to the one observed for droplets. 
Correspondingly, the elastic regime bears strong resemblance to the cases of highly viscous drops \cite{langley:2017} at low impact speeds.
In the latter case, a measure of the viscous pressure in the drop is given by $P_{visc} \sim \mu_i V / \mathcal{L}$, where $\mu_i$ is the viscosity of the impacting droplet.
Balancing it with the lubrication pressure, we obtain 
\begin{align}
&\mathcal{L}_{visc} = R \, (\mu_m' / \mu_i)^{1/3} \, ,
&\mathcal{H}_{visc} = R \, (\mu_m' / \mu_i)^{2/3} \, ,
\end{align}
while the inertial scaling is identical to the one presented for the solid impactor in \cref{eq:inertial_scales}. 
By imposing $P_{visc} \sim P_{\text{in}}$ and following the same line of reasoning presented in sections \ref{sec:scaling_transition} and \ref{sec:scaling_fluid_compressibility}, we obtain the corresponding transition parameter:
\begin{align}
    \Phi_{visc \to in} =
    \Phi_{in \to visc} =
    \Phi 
    &=
    {\mathcal{L}_{\text{in}}^2 / \eta_i \over H_{\text{in}}/V}
     \nonumber \\
    &={\rho_i V R \over \mu_i} 
    = 
    {\tau_{\text{propagation}} 
        \over 
    \tau_{\text{impact}}} 
    =
    \mathrm{Re}_i
    \, ,
    \label{eq:transition_parameter_droplets}
\end{align}
which is the Reynolds number of the droplet ($\eta_i$ represents its kinematic viscosity). 
Again, this can be interpreted as the ratio of the time scales of the phenomenon, with the propagation time now depending on the viscous diffusion timescale ($\mathcal{L}_{\text{in}}^2 / \eta_i $) \cite{batchelor:2000} instead of the wave propagation speed.
% Reminiscent of boundary layer thickness
%
Due to the parabolic geometry of the tip, the deformation front again expands proportionally to $\sim t^{1/2}$, which in this case corresponds to the power law of signal propagation, i.e. viscous diffusion.  
This suggests that catching-up of viscous stresses with the deformation front could only occur if the two phenomena arise at different times. 
If this is not the case, the dimple radius should vary smoothly when transitioning between the two limit regimes.
Experimental evidence corroborating these hypotheses can be found in \cite{langley:2017}. For example in their Figure 7 \cite{langley:2017}, effects of viscosity become noticeable when $\mathrm{Re}_i \approx 2$ and all profiles appear smooth, contrarily to \Cref{fig:radius_time} at the transition.

Conversely, if surface tension, $\gamma$, dominates over viscosity, the pressure scale is given by $P_{\text{surf}} \sim \gamma \kappa$, where $\kappa$ is the interface curvature. 
From the dimple geometry we can estimate $\kappa \sim H/\mathcal{L}^2$ and balancing this pressure with the lubrication one we get \cite{Bouwhuis:2012}:
\begin{align}
&\mathcal{L}_{surf} = R \, (\mu_m' V / \gamma)^{1/4} \, ,
&\mathcal{H}_{surf} = R \, (\mu_m' V/ \gamma)^{1/2} \, .
\end{align}
Introducing the capillary number $\mathrm{Ca} = \mu'_m V / \gamma$, following the above reasoning once again, the crossover when capillary effects and inertia become of the same order is governed by $\Psi_{in \to surf}^{3/4} = \mathrm{Stk} \, \mathrm{Ca}^{3/4}$, the same parameter as in \cite{Bouwhuis:2012}. This dimensionless groups can also be formed with the Weber number in lieu of the capillary number \cite{Ruiter:2012}.
By computing $\Psi_{surf \to in}$, we notice that: 
\begin{align}
    &\Psi_{surf \to in}^{2/3} =
    \Psi_{in \to surf}^{1/2}
    =
    \Psi 
    =
    {\mathcal{L}_{\text{in}}/ c_{cap} \over H_{\text{in}}/V}
     \nonumber \\
    &=\left({\rho_i^4 V^7 R^4 \over \gamma^3 \mu'_m}\right)^{1/6} 
    = 
    {\tau_{\text{propagation}} 
        \over 
    \tau_{\text{impact}}}
    \, ,
\label{eq:transition_parameter_droplets_capillar}
\end{align}
where $\tau_{\text{propagation}}$ in this case depends on the capillary waves' velocity $c_{cap}$.

We assumed a deep-water capillary wave, so the characteristic value of velocities $c_{cap} = (\gamma \omega / \rho_i )^{1/3}$ \cite{landau:2013}, 
plus the characteristic frequency of the waves $\omega$ being the inverse of the characteristic time, which entails $c_{cap} = V^{1/6} \gamma^{1/2} R^{-1/3} \mu'^{-1/6} \rho_i^{-1/3}$. 
In turn, this implies a characteristic wavelength $\Lambda = R^{2/3} \mu'^{1/3} V^{-1/3} \rho_i^{-1/3}$, which yields values consistent with those measured experimentally post-impact \cite{li:2019}. 
Finally, we speculate that the ``double dimple'' and the jump observed numerically and experimentally in the trapped bubble volume \cite{Bouwhuis:2012} as a function of $\gamma$ can be explained within the framework that we developed in Section \ref{sec:maximum_bubble}, by replacing $\phi$ with $\Psi$ and the solid's Rayleigh waves with the droplet's capillary ones. 
Specifically, the finding in \cite{Bouwhuis:2012} that the maximal air entrapment bubble happens when $\mathrm{Stk} \, \mathrm{Ca}^{3/4} \sim 1$ is equivalent to saying that it happens when $\tau_{\text{impact}} \sim \tau_{\text{propagation}}$, i.e., similar to other scenarios discussed above, when signals in the impactor (capillary waves in this case) can catch up with the pressure build-up, touchdown is delayed and the bubble has more time to grow. 
\textcolor{black}{Both} the eventual appearance of a ``\textit{double kink}'' described in \cite{Ruiter:2012} (for low impact velocities) \textcolor{black}{and the increased ``\textit{time-to-contact}'' associated to the ``\textit{skating impact mode}'' \cite{Sprittles:2023}} could also be a manifestation of the same phenomenon.

\section{Summary and outlook}
\label{sec:final}

The work unveiled, via numerical simulations and scaling analysis, the mechanisms underlying the fluid-mediated dynamics of soft solids approaching a rigid smooth surface. 
By matching the numerical simulations with experimental data for the dimple radius, we verified the validity of scaling arguments previously proposed and extended them. 
Additionally, the critical parameter $\phi$ determining the cross-over between elastic and inertial regime was identified as the ratio of the time it takes for the lubrication to locally slow down the impactor, to the time it takes for information about the impending impact to propagate in the solid. 
We showed how a similar reasoning can help to quantify the influence of solid compressibility during impact, 
which adds another layer of complexity compared to the fluid droplet case and another dimensionless parameter to consider, $\psi$. 
Our theory predicts and characterizes the elastic and inertial regimes.
It provides an explanation for the presence of a maximum in the entrapped air bubble radius, as seen in both experiments and simulations.
Moreover, it also highlights that the inertia-dominated regime is inextricably linked to either impactor's or mediating fluid's compressibility, or with both simultaneously. 
The competition of time scales may also shine new light on formerly observed phenomena in highly viscous drops.
We also offered an explanation for the abrupt reduction in the lateral scale of the bubble, which occurs when inertial forces start to dominate the impactor response, attributed to the parabolic geometry of the impactor tip.

Several questions regarding the role of compressibility remain unanswered. 
For instance, if lubricating fluid layer is less compressible than the solid, should we anticipate an impact event without any physical contact, across all impact velocities? 
Furthermore, the analysis here has focused on the canonical hemispherical impactor geometry; however, in general solids retain their shape, and often take on complex geometries that will modify the lubricating fluid flow and resulting lubrication stresses. 
This begs the question - beyond the radius of curvature of the impactor, what is the role of the impactor geometry in establishing the lubricating stress and entrainment? 
And could this be exploited, for instance in applications where large contact areas are desirable, as in end effectors for gripping material? 
Our analysis highlights the critical role played by the relative time scales of Rayleigh wave propagation and the forced deformation front velocity of the impactor. 
If the impactor is viscoelastic, is there an analagous time scale that one might envision that plays a similar role in governing the contact process? 
With our study, we have identified a sharp transition between elasticity-dominated impacts and inertially-dominated impacts. 
How this transition manifests itself in the myriad physical scenarios where soft objects are in contact may be an essential step toward understanding the strongly nonlinear dynamics observed in suspensions and other dense collections of particles, including biological cells.

%%%

\section*{Acknowledgements}

This project has received funding from the European Union’s Horizon 2020 research and innovation programme under the Marie Skłodowska-Curie grant agreement No 945363.

\section*{Author credit}

J.M.K. proposed the original idea of soft matter impact. 
%
%J. G.-S. introduced the concept of ratio of timescales. 
%
J.B., B.L. and J.G.-S. developed the scaling analysis. 
J.B. and J.G.-S. planned the numerical study. 
J.B. performed the simulations. 
All authors discussed, interpreted the results, and contributed to writing the manuscript.\\

\appendix
\section{Simulations}
\label{subsec:simulations}
We model the interaction between the impacting solid particle and the fluid surrounding it by means of monolithic  finite element simulations for the fluid-structure interaction problem.
To simplify the computation, we study the axisymmetric case as we are primarily interested in normal impacts.
The schematic of the simulation is illustrated in \Cref{fig:impact_configuration}.
We represent the relevant leading boundary of the impacting sphere with $\partial \Omega^-$.  

The initial gap, $h_0 \approx 12\mu m$, is set such that no considerable deformation takes place in the early stages of the simulation (see that the elasticity parameter \cite{David:1986} $\approx 0.004-0.35$, where the highest value is obtained in the case $V=5 \, m/s$ for the soft impactor).
At the same time, in order to neglect inertial effects in the thin film, we require \cite{David:1986} $\rho_i V h_0/ \mu_m \ll 1$, therefore we should not increase $h_0$ significantly. 
Simulations with the same time step showed limited discrepancy with values of the initial gap $\approx 6\mu m, \, 30 \mu m $ (see \Cref{fig:r_sim_exp}), except near the transition where the stress singularity stretches the capability of the numerical method.
We observed gliding up to higher values of velocity when reducing the initial gap, signaling that the value of $\phi$ when the sharp drop occurs is sensitive to the initial conditions, i.e. the instance when velocity is measured before impact. 
A slightly higher velocity, caused by the film having less time to slow down the impactor (smaller $h_0$), could cause pressure spikes and in turn gliding. 

For the solid mechanics we employ the Neo-Hookean model, which has been widely used to model polymers and rubber-like materials undergoing large deformations. 
In this setting, $\vecsym{x} = (r, \theta, z)$ are the spatial coordinates, 
$\vecsym{u} = (u_r(r,z),0, u_z(r,z))$ is the displacement field and the deformation gradient is defined as $
    \vecsym{F} 
    = 
    %{\partial \vecsym{\mathrm{x}} 
    %\over 
    %\partial \vecsym{\mathrm{X}}} 
    %= 
    \vecsym{I} 
    + 
    \nabla    
    % _{\vecsym{X}} 
    \vecsym{u}$.

Consequently, the Green-Lagrange tensor is written as \cite{Mardsen:1994}:
\begin{align}
    \vecsym{E} = 
    {1 \over 2}
    \left(
    \vecsym{F}^\top \vecsym{F} - \vecsym{I}
    \right) \, .
\end{align}
The corresponding work-conjugate is the second Piola-Kirchoff stress defined as:
\begin{align}
    \vecsym{S} = 
    J \vecsym{F}^{-1} \vecsym{\sigma} \vecsym{F}^{-\top} \, ,
\end{align}
where $J = \det (\vecsym{F})$ and $\vecsym{\sigma}$ is the Cauchy stress tensor. 
The Neo-Hookean strain energy density is defined by
\begin{align}
    \psi(\vecsym{E}) = G \left(\mathrm{tr}(\vecsym{E}) - \ln J \right) + {\lambda \over 2} \left(\ln J \right)^2 \, ,
\end{align}
where  $\lambda, G$ are the Lam\'e constants of the impactor material. 
These are related to the Young's modulus $E$ and the Poisson's ratio $\nu$, via the usual relations:
\begin{align}
    &E = {
    G (3\lambda + 2 G) 
    \over 
    \lambda + G} \, , 
    &\nu 
    = {\lambda \over 2 (\lambda + G)}
\end{align}
The Neo-Hookean hyperelastic constitutive relationship is obtained by deriving the strain energy density with respect to the Green-Lagrange strain
\begin{align}
    \vecsym{S} = {\partial \psi \over \partial \vecsym{E}}   \, . 
\end{align}
In absence of volume forces, denoting by $(\cdot)^T$ the transpose operator, linear momentum balance equations in the reference configuration for a solid undergoing finite deformations are:
\begin{align}
    \rho_0
    {\partial^2 \vecsym{u} 
    \over 
    \partial t^2} 
    = 
    \nabla
    %_{\vecsym{X}} 
    \cdot (\vecsym{F}\vecsym{S})^T \, ,
    \label{eq:navier_finite}
\end{align}
where $\rho_0$ is the density in the initial configuration. 

In the cushioning fluid, we use lubrication equation, provided $\text{Stk}^{-1/3} \gg \rho_m/\rho_i$.
Thus, in the thin film the vertical gap is so narrow that average fields are considered a good approximation.
In the incompressible axisymmetric form \cite{szeri:2010}, the Reynolds equation is written in \Cref{eq:reynolds}.
The boundary condition for the impactor on the leading edge is \cref{eq:bc}.
Finally, the initial pressure profile for the thin film is set to zero, effectively neglecting the contribution of ambient pressure, and the boundary condition $p_m(r=R) = 0$ is enforced.

An implicit backward differentiation scheme is used, with at least second order accuracy.
A damped Newton method algorithm is adopted, with minimum relative tolerance of $10^{-5}$ on the weighted residual. 
Both the solid and fluid mechanics problem are solved in finite element formulation with the \textit{Comsol Multiphysics} \textregistered \hspace{.1mm} software, in a fully-coupled fashion, similarly to what is done in \cite{habchi:2008}.
Quadratic elements are used for both the fluid flow and the structural mechanics.
Mesh sizes ranging from $1 \mu m$ to $100 \mu m$ on the region where contact is expected to take place, give similar results when $\phi \gg 1$ and $\phi \ll 1$.
The same occurs for the time step size: defining $T=h_0/V$ the time it would take for the impactor to make contact if no deformation was present, time steps of the order $T/10$ and $ T/1000$ yield similar results in the slow and fast impact limits.
However, in the instance of $\phi \sim 1$ the system comes close to a singularity and then moves away from it as was shown in \cref{fig:impact_phi_order_1}.
Therefore, finer meshes  and smaller time steps give nonphysical numerical instabilities.
The final choice is to adopt a mesh size $20 \mu m$ and a time step $\Delta t = T/15$.
 
The simulations are terminated when:
\begin{itemize}
    \item the minimum gap reaches values of the order of $h_{min} \approx 100 \, nm$, as the local Knudsen number (the ratio of the film thickness to the mean free path of molecules) is of the order of unity and molecular flow should be taken into consideration \cite{Mandre:2010,batchelor:2000}.
    The last converged time step before this occurrence is taken as $t_{\text{final}}$
    \item the solver is unable to converge as a singularity of stress is reached. 
    This is the case for the very low velocity impacts and when $\mathcal{L}$ drops suddenly transitioning from the elastic to the inertial regime (see \Cref{tab:final_gap}).
    We identify the time when this happens as $t_{\text{final}}$.
\end{itemize}
Therefore, bubble radii for some simulations are deemed to be underestimated. \Cref{eq:geometrical_law} can be used to give an estimate of the error, yielding a maximum value $\approx 7 \mu m$, small compared to $\mathcal{L}$.

\begin{table}[h]
    \centering
    \caption{Values of the final height for the last time-step in output for simulations in \Cref{fig:r_sim_exp}}
    \begin{ruledtabular}
        \begin{tabular}{ccc}
            Velocity $[m/s]$  & ``Soft''$[\mu m]$   & ``Hard'' $[\mu m]$\\ \hline
            0.05 & 0.51 & 0.25\\ 
            0.10 & 0.38 & 0.22 \\ 
            0.25 & 0.12 & 0.11 \\ 
            0.40 & 0.10 & 0.12\\ 
            0.49 & 0.13 & 0.11 \\
            0.74 & 0.11 & 0.12  \\
            0.99 & 0.12 & 0.13   \\
            1.24 & 0.54 & 0.11 \\ 
            1.43 & 0.38 & 0.12  \\ 
            1.60 & 0.28 & 0.12 \\ 
            2.03 & 0.22 & 0.44 \\ 
            2.43 & 0.19 & 0.37 \\ 
            3.38 & 0.11 & 0.24\\ 
            4.53 & 0.11 & 0.13 \\
            5.42 & 0.13 & 0.14\\
        \end{tabular}    
    \end{ruledtabular}
    \label{tab:final_gap}
\end{table}

\section{Dimensionless Groups}
\label{sec:app_dimensionless_groups}
The scaling presented in Section \ref{sec:scaling} can be alternatively obtained by identifying the governing dimensionless groups of the physical phenomenon.

The solid momentum balance \Cref{eq:navier_finite}, in the small strain limit, reduces to:
\begin{align}
    \rho_i
    {\partial \vecsym{v} 
    \over 
    \partial t} 
    = 
    \nabla
    %_{\vecsym{X}} 
    \cdot
    \vecsym{\sigma} \, ,
\label{eq:navier_small}
\end{align}
where $\vecsym{v} = \partial \vecsym{u} / \partial \vecsym{t}$ and the Cauchy stress is given by:
\begin{align}
    \vecsym{\sigma} = 2 G \bm{\varepsilon}
    +\left(K-{2 \over 3} G\right) tr(\bm{\varepsilon}) \vecsym{I} \, .
    \label{eq:constitutive_stress}
\end{align}
The fluid mechanics for the thin film is described by \Cref{eq:reynolds}.
The ideal gas state equation for a polytropic process reads: 
\begin{align}
    {p \over p_0} = 
    \left({\rho_m \over \rho_{m,0}} \right)^{\upgamma} \, .
    \label{eq:gas_state}
\end{align}
with a certain constant $\upgamma$ which depends on the specific process.
Finally, from \Cref{eq:parabolic_approx} the appearance of a dimple at location $r=r^*(t)$ is obtained by setting $\partial h(r,t) / \partial r = 0$ leading to:
\begin{align}
    {r^*(t) \over R} + {\partial w \over \partial r} = 0 \, ,
    \label{eq:dimple}
\end{align}
which is where the geometry of the problem enters.

We can perform dimensional analysis by appropriately scaling the physical dimensions around the tip of the impactor, where most of the dynamics is supposed to take place (see \Cref{fig:impact_configuration}).
For lubrication equations to be valid we require $H/\mathcal{L} \ll 1$, i.e. the height of the gap to be much smaller than its width.
This results in the radial velocity in the fluid to be much greater than the vertical one, to satisfy mass conservation.
We also assume that the gap height is set by the scale of solid vertical deformation $w(r,t)$.
The resulting scaling is: 
\begin{subequations}
\begin{align}
    t &=\tau \tilde{t} \\
    \vecsym{v} &=V \tilde{\vecsym{v}} \\
    \left[r, r^*(t) \right]
    &=
    \mathcal{L} \left[\tilde{r},\tilde{r}^*(\tilde{t})\right] \, , \\
    \left[ h, w \right]
    &=
    H \left[\tilde{h},\tilde{w} \right] \, , \\
    p(r,t) 
    &= P \tilde{p}(\tilde{r}, \tilde{t}) \\
    \vecsym{\sigma}(r,w,t)
    &=
    P \tilde{\vecsym{\sigma}}(\tilde{r}, \tilde{w}, \tilde{t}) \, ,
\end{align}
\end{subequations}
where we set the same pressure for the fluid and the solid stress.

We proceed by plugging in the rescaled quantities in the equations formerly discussed.
\Cref{eq:navier_small} yields:
\begin{align}
    \mathcal{G}_i {\partial \tilde{\vecsym{v}} \over \partial \tilde{t}} = 
    \tilde{\nabla} \cdot \tilde{\vecsym{\sigma}} \, ,
    && \mathcal{G}_i = {\rho_i V\mathcal{L}\over \tau P}
    \label{eq:dimensionless_inertial_group}
\end{align}
which is the ratio of the solid inertia pressure scale over the pressure scale of the problem.
From \Cref{eq:reynolds}, we get:
\begin{align}
    \mathcal{G}_m {\partial (\tilde{p}^{1/\upgamma} \tilde{h}) \over \partial \tilde{t}} =
    {1 \over \tilde{r}}
    \left(
    \tilde{r} \tilde{p}^{1/\upgamma} \tilde{h}^3 
    {\partial \tilde{p} \over \partial \tilde{r}}
    \right) \, ,
    && \mathcal{G}_m = {12 \mu_m' \mathcal{L}^2 \over P \tau H^2} \,,
    %\label{eq:dimensionless_inertial_group}
\end{align}
the lubrication dimensionless group.
The dimple geometry, \Cref{eq:geometrical_law} leads to:
\begin{align}
    \mathcal{G}_d \tilde{r}^* + {\partial \tilde{w}\over \partial \tilde{r}} = 0 \, ,
    && \mathcal{G}_d = {\mathcal{L}^2 \over R H} \,.
    %\label{eq:dimensionless_inertial_group}
\end{align}
In \Cref{eq:constitutive_stress} there are several possibilities to gauge the value of strain.
If we assume the solid to deform over the scale of the dimple, strain can be scaled as $\bm{\varepsilon}=(H/\mathcal{L})\tilde{\bm{\varepsilon}}$, which leads to:
\begin{align}
    \tilde{\vecsym{\sigma}} = \mathcal{G}_e 
    \left(
    2  \tilde{\bm{\varepsilon}} +
    \left({K\over G} - {2 \over 3}\right)
    tr(\tilde{\bm{\varepsilon}})
    \vecsym{I}
    \right)
    \, ,
     && \mathcal{G}_e = {G H  \over P \mathcal{L}} \,,
    \label{eq:dimensionless_elastic_group}
\end{align}
Alternatively, for stresses manifesting where pressure waves propagate we can write $\bm{\varepsilon}={H \over c_p \tau} \tilde{\bm{\varepsilon}}$ and: 
\begin{align}
    \tilde{\vecsym{\sigma}}
    \approx
    \mathcal{G}_M 
    \tilde{\bm{\varepsilon}}
    \, ,
     && \mathcal{G}_M = {H M\over c_p \tau P } \,,
    %\label{eq:dimensionless_elastic_group}
\end{align}
where we assumed an uniaxial strain scenario.
Additionally, gas compressibility effects (\Cref{eq:gas_state}) are described by the dimensionless group $\mathcal{G}_c = (P/p_0)^{1/\upgamma}$
and we set the time scale as the one governing lubrication flow
$\tau = H/V$.
Assuming dimple formation and thin film flow are in effect corresponds to imposing $\mathcal{G}_d =1 \implies\mathcal{L}= \sqrt{R H}$, and $\mathcal{G}_m =1$.
The elastic, inertial and solid-compressible regimes are obtained by enforcing $\mathcal{G}_e$, $\mathcal{G}_i$ and $\mathcal{G}_M$ being of order one respectively.
By doing so we obtain the values of $\mathcal{L}, \, H, \,\tau, \,P$ in each regime and we can substitute them in the dimensionless groups to asses their relative importance.

An analogous way of reasoning can be used to scale the droplet impact.
The full incompressible Navier-Stokes equation is:
\begin{align}
    \rho_i
    {\partial \vecsym{v} 
    \over 
    \partial t} 
    = 
    \nabla
    %_{\vecsym{X}} 
    \cdot
    \vecsym{\sigma}
    -\rho_i 
    (\vecsym{v}\cdot \nabla \vecsym{v})  \, ,
\label{eq:navier_full}
\end{align}
when we used the Newtonian stress for the fluid:
\begin{align}
    \vecsym{\sigma} =
    \- p_i \vecsym{I} 
    + \mu_i 
    (\nabla \vecsym{v} + \nabla \vecsym{v}^T ) \, .
    \label{eq:constitutive_newtonian}
\end{align}
The normal boundary condition at the droplet-gas interface is
\begin{align}
    (p-\vecsym{\sigma} \cdot \vecsym{n})= \gamma \kappa \, ,
    \label{eq:bc_capillary_viscous}
\end{align}
where the mean curvature $\kappa$ can be approximated as in Section \ref{sec:droplet_analysis}.
The gas state equation and the dimple condition remain unchanged.

By performing the same scaling we obtain from \Cref{eq:navier_full}:
\begin{align}
    \mathcal{G}_i {\partial \tilde{\vecsym{v}} \over \partial \tilde{t}} = 
    \tilde{\nabla} \cdot \tilde{\vecsym{\sigma}}
    - {\rho_i V^2 \over P} \, (\tilde{\vecsym{v}} \cdot \tilde{\nabla} \tilde{\vecsym{v}}),
    && \mathcal{G}_i = {\rho_i V\mathcal{L}\over \tau P}
    \label{eq:dimensionless_inertial_group_droplet}
\end{align}
and $(\rho_i V^2/P)/\mathcal{G}_i = H/\mathcal{L} \ll 1$ which entails the non-linear term can be neglected.
\Cref{eq:bc_capillary_viscous} yields:
\begin{align}
    \tilde{p}& -
    {p_i \over P} (\vecsym{I} \cdot \vecsym{n})
    + \mathcal{G}_v
    ((\tilde{\nabla} \tilde{\vecsym{v}} + \tilde{\nabla} \tilde{\vecsym{v}}^T ) \cdot &\vecsym{n})
    = \mathcal{G}_s {\tilde{h} \over \tilde{r}^{*2} }, \\
    \mathcal{G}_v &= {\mu_i V \over \mathcal{L} P}, 
    &\mathcal{G}_s = {\gamma H \over \mathcal{L}^2 P}
    \label{eq:dimensionless_viscosity_group}
\end{align}
where we defined the inertial and the capillary dimensionless groups.
Assuming $p_i/P \ll 1$,  the viscous, inertial and capillary regimes are obtained by enforcing $\mathcal{G}_v$, $\mathcal{G}_i$ and $\mathcal{G}_s$ being of order one respectively.

\bibliography{references}% Produces the bibliography via BibTeX.

%apsrev4-2.bst 2019-01-14 (MD) hand-edited version of apsrev4-1.bst
%Control: key (0)
%Control: author (8) initials jnrlst
%Control: editor formatted (1) identically to author
%Control: production of article title (0) allowed
%Control: page (0) single
%Control: year (1) truncated
%Control: production of eprint (0) enabled
\providecommand{\noopsort}[1]{}\providecommand{\singleletter}[1]{#1}%
\begin{thebibliography}{39}%
\makeatletter
\providecommand \@ifxundefined [1]{%
 \@ifx{#1\undefined}
}%
\providecommand \@ifnum [1]{%
 \ifnum #1\expandafter \@firstoftwo
 \else \expandafter \@secondoftwo
 \fi
}%
\providecommand \@ifx [1]{%
 \ifx #1\expandafter \@firstoftwo
 \else \expandafter \@secondoftwo
 \fi
}%
\providecommand \natexlab [1]{#1}%
\providecommand \enquote  [1]{``#1''}%
\providecommand \bibnamefont  [1]{#1}%
\providecommand \bibfnamefont [1]{#1}%
\providecommand \citenamefont [1]{#1}%
\providecommand \href@noop [0]{\@secondoftwo}%
\providecommand \href [0]{\begingroup \@sanitize@url \@href}%
\providecommand \@href[1]{\@@startlink{#1}\@@href}%
\providecommand \@@href[1]{\endgroup#1\@@endlink}%
\providecommand \@sanitize@url [0]{\catcode `\\12\catcode `\$12\catcode `\&12\catcode `\#12\catcode `\^12\catcode `\_12\catcode `\%12\relax}%
\providecommand \@@startlink[1]{}%
\providecommand \@@endlink[0]{}%
\providecommand \url  [0]{\begingroup\@sanitize@url \@url }%
\providecommand \@url [1]{\endgroup\@href {#1}{\urlprefix }}%
\providecommand \urlprefix  [0]{URL }%
\providecommand \Eprint [0]{\href }%
\providecommand \doibase [0]{https://doi.org/}%
\providecommand \selectlanguage [0]{\@gobble}%
\providecommand \bibinfo  [0]{\@secondoftwo}%
\providecommand \bibfield  [0]{\@secondoftwo}%
\providecommand \translation [1]{[#1]}%
\providecommand \BibitemOpen [0]{}%
\providecommand \bibitemStop [0]{}%
\providecommand \bibitemNoStop [0]{.\EOS\space}%
\providecommand \EOS [0]{\spacefactor3000\relax}%
\providecommand \BibitemShut  [1]{\csname bibitem#1\endcsname}%
\let\auto@bib@innerbib\@empty
%</preamble>
\bibitem [{\citenamefont {Rallabandi}(2024)}]{Rallabandi:2023}%
  \BibitemOpen
  \bibfield  {author} {\bibinfo {author} {\bibfnamefont {B.}~\bibnamefont {Rallabandi}},\ }\bibfield  {title} {\bibinfo {title} {Fluid-elastic interactions near contact at low reynolds number},\ }\bibfield  {journal} {\bibinfo  {journal} {Annual Review of Fluid Mechanics}\ }\textbf {\bibinfo {volume} {56}},\ \href {https://doi.org/10.1146/annurev-fluid-120720-024426} {10.1146/annurev-fluid-120720-024426} (\bibinfo {year} {2024}),\ \Eprint {https://arxiv.org/abs/https://doi.org/10.1146/annurev-fluid-120720-024426} {https://doi.org/10.1146/annurev-fluid-120720-024426} \BibitemShut {NoStop}%
\bibitem [{\citenamefont {Zheng}\ \emph {et~al.}(2021)\citenamefont {Zheng}, \citenamefont {Dillavou},\ and\ \citenamefont {Kolinski}}]{Zheng:2021}%
  \BibitemOpen
  \bibfield  {author} {\bibinfo {author} {\bibfnamefont {S.}~\bibnamefont {Zheng}}, \bibinfo {author} {\bibfnamefont {S.}~\bibnamefont {Dillavou}},\ and\ \bibinfo {author} {\bibfnamefont {J.~M.}\ \bibnamefont {Kolinski}},\ }\bibfield  {title} {\bibinfo {title} {Air mediates the impact of a compliant hemisphere on a rigid smooth surface},\ }\href@noop {} {\bibfield  {journal} {\bibinfo  {journal} {Soft Matter}\ }\textbf {\bibinfo {volume} {17}},\ \bibinfo {pages} {3813} (\bibinfo {year} {2021})}\BibitemShut {NoStop}%
\bibitem [{\citenamefont {Xue}\ \emph {et~al.}(2021)\citenamefont {Xue}, \citenamefont {Gu}, \citenamefont {Li}, \citenamefont {Yu}, \citenamefont {Yin}, \citenamefont {Qin}, \citenamefont {Jiang}, \citenamefont {Wang},\ and\ \citenamefont {Cao}}]{xue:2021}%
  \BibitemOpen
  \bibfield  {author} {\bibinfo {author} {\bibfnamefont {B.}~\bibnamefont {Xue}}, \bibinfo {author} {\bibfnamefont {J.}~\bibnamefont {Gu}}, \bibinfo {author} {\bibfnamefont {L.}~\bibnamefont {Li}}, \bibinfo {author} {\bibfnamefont {W.}~\bibnamefont {Yu}}, \bibinfo {author} {\bibfnamefont {S.}~\bibnamefont {Yin}}, \bibinfo {author} {\bibfnamefont {M.}~\bibnamefont {Qin}}, \bibinfo {author} {\bibfnamefont {Q.}~\bibnamefont {Jiang}}, \bibinfo {author} {\bibfnamefont {W.}~\bibnamefont {Wang}},\ and\ \bibinfo {author} {\bibfnamefont {Y.}~\bibnamefont {Cao}},\ }\bibfield  {title} {\bibinfo {title} {Hydrogel tapes for fault-tolerant strong wet adhesion},\ }\href@noop {} {\bibfield  {journal} {\bibinfo  {journal} {Nature Communications}\ }\textbf {\bibinfo {volume} {12}},\ \bibinfo {pages} {7156} (\bibinfo {year} {2021})}\BibitemShut {NoStop}%
\bibitem [{\citenamefont {Higgins}\ and\ \citenamefont {Mahadevan}(2010)}]{higgins:2010}%
  \BibitemOpen
  \bibfield  {author} {\bibinfo {author} {\bibfnamefont {J.~M.}\ \bibnamefont {Higgins}}\ and\ \bibinfo {author} {\bibfnamefont {L.}~\bibnamefont {Mahadevan}},\ }\bibfield  {title} {\bibinfo {title} {Physiological and pathological population dynamics of circulating human red blood cells},\ }\href@noop {} {\bibfield  {journal} {\bibinfo  {journal} {Proceedings of the National Academy of Sciences}\ }\textbf {\bibinfo {volume} {107}},\ \bibinfo {pages} {20587} (\bibinfo {year} {2010})}\BibitemShut {NoStop}%
\bibitem [{\citenamefont {Coussot}\ and\ \citenamefont {Ancey}(1999)}]{coussot:1999}%
  \BibitemOpen
  \bibfield  {author} {\bibinfo {author} {\bibfnamefont {P.}~\bibnamefont {Coussot}}\ and\ \bibinfo {author} {\bibfnamefont {C.}~\bibnamefont {Ancey}},\ }\bibfield  {title} {\bibinfo {title} {Rheophysical classification of concentrated suspensions and granular pastes},\ }\href@noop {} {\bibfield  {journal} {\bibinfo  {journal} {Phys. Rev. E}\ }\textbf {\bibinfo {volume} {59}},\ \bibinfo {pages} {4445} (\bibinfo {year} {1999})}\BibitemShut {NoStop}%
\bibitem [{\citenamefont {Mandre}\ \emph {et~al.}(2009)\citenamefont {Mandre}, \citenamefont {Mani},\ and\ \citenamefont {Brenner}}]{Mandre:2009}%
  \BibitemOpen
  \bibfield  {author} {\bibinfo {author} {\bibfnamefont {S.}~\bibnamefont {Mandre}}, \bibinfo {author} {\bibfnamefont {M.}~\bibnamefont {Mani}},\ and\ \bibinfo {author} {\bibfnamefont {M.~P.}\ \bibnamefont {Brenner}},\ }\bibfield  {title} {\bibinfo {title} {Precursors to splashing of liquid droplets on a solid surface},\ }\href {https://doi.org/10.1103/PhysRevLett.102.134502} {\bibfield  {journal} {\bibinfo  {journal} {Phys. Rev. Lett.}\ }\textbf {\bibinfo {volume} {102}},\ \bibinfo {pages} {134502} (\bibinfo {year} {2009})}\BibitemShut {NoStop}%
\bibitem [{\citenamefont {Mani}\ \emph {et~al.}(2010)\citenamefont {Mani}, \citenamefont {Mandre},\ and\ \citenamefont {Brenner}}]{Mandre:2010}%
  \BibitemOpen
  \bibfield  {author} {\bibinfo {author} {\bibfnamefont {M.}~\bibnamefont {Mani}}, \bibinfo {author} {\bibfnamefont {S.}~\bibnamefont {Mandre}},\ and\ \bibinfo {author} {\bibfnamefont {M.~P.}\ \bibnamefont {Brenner}},\ }\bibfield  {title} {\bibinfo {title} {Events before droplet splashing on a solid surface},\ }\href {https://doi.org/10.1017/S0022112009993594} {\bibfield  {journal} {\bibinfo  {journal} {Journal of Fluid Mechanics}\ }\textbf {\bibinfo {volume} {647}},\ \bibinfo {pages} {163–185} (\bibinfo {year} {2010})}\BibitemShut {NoStop}%
\bibitem [{\citenamefont {Kolinski}\ \emph {et~al.}(2012)\citenamefont {Kolinski}, \citenamefont {Rubinstein}, \citenamefont {Mandre}, \citenamefont {Brenner}, \citenamefont {Weitz},\ and\ \citenamefont {Mahadevan}}]{kolinski:2012}%
  \BibitemOpen
  \bibfield  {author} {\bibinfo {author} {\bibfnamefont {J.~M.}\ \bibnamefont {Kolinski}}, \bibinfo {author} {\bibfnamefont {S.~M.}\ \bibnamefont {Rubinstein}}, \bibinfo {author} {\bibfnamefont {S.}~\bibnamefont {Mandre}}, \bibinfo {author} {\bibfnamefont {M.~P.}\ \bibnamefont {Brenner}}, \bibinfo {author} {\bibfnamefont {D.~A.}\ \bibnamefont {Weitz}},\ and\ \bibinfo {author} {\bibfnamefont {L.}~\bibnamefont {Mahadevan}},\ }\bibfield  {title} {\bibinfo {title} {Skating on a film of air: drops impacting on a surface},\ }\href@noop {} {\bibfield  {journal} {\bibinfo  {journal} {Phys. Rev. Lett.}\ }\textbf {\bibinfo {volume} {108}},\ \bibinfo {pages} {074503} (\bibinfo {year} {2012})}\BibitemShut {NoStop}%
\bibitem [{\citenamefont {Josserand}\ and\ \citenamefont {Thoroddsen}(2016)}]{Thoroddsen:2016}%
  \BibitemOpen
  \bibfield  {author} {\bibinfo {author} {\bibfnamefont {C.}~\bibnamefont {Josserand}}\ and\ \bibinfo {author} {\bibfnamefont {S.}~\bibnamefont {Thoroddsen}},\ }\bibfield  {title} {\bibinfo {title} {Drop impact on a solid surface},\ }\href {https://doi.org/10.1146/annurev-fluid-122414-034401} {\bibfield  {journal} {\bibinfo  {journal} {Annual Review of Fluid Mechanics}\ }\textbf {\bibinfo {volume} {48}},\ \bibinfo {pages} {365} (\bibinfo {year} {2016})},\ \Eprint {https://arxiv.org/abs/https://doi.org/10.1146/annurev-fluid-122414-034401} {https://doi.org/10.1146/annurev-fluid-122414-034401} \BibitemShut {NoStop}%
\bibitem [{\citenamefont {Wu}\ \emph {et~al.}(2021)\citenamefont {Wu}, \citenamefont {Cao},\ and\ \citenamefont {Yao}}]{Wu:2021}%
  \BibitemOpen
  \bibfield  {author} {\bibinfo {author} {\bibfnamefont {Z.}~\bibnamefont {Wu}}, \bibinfo {author} {\bibfnamefont {Y.}~\bibnamefont {Cao}},\ and\ \bibinfo {author} {\bibfnamefont {Y.}~\bibnamefont {Yao}},\ }\bibfield  {title} {\bibinfo {title} {Anatomy of air entrapment in drop impact on a solid surface},\ }\href {https://doi.org/https://doi.org/10.1016/j.ijmultiphaseflow.2021.103724} {\bibfield  {journal} {\bibinfo  {journal} {International Journal of Multiphase Flow}\ }\textbf {\bibinfo {volume} {142}},\ \bibinfo {pages} {103724} (\bibinfo {year} {2021})}\BibitemShut {NoStop}%
\bibitem [{\citenamefont {Riboux}\ and\ \citenamefont {Gordillo}(2014)}]{riboux:2014}%
  \BibitemOpen
  \bibfield  {author} {\bibinfo {author} {\bibfnamefont {G.}~\bibnamefont {Riboux}}\ and\ \bibinfo {author} {\bibfnamefont {J.~M.}\ \bibnamefont {Gordillo}},\ }\bibfield  {title} {\bibinfo {title} {Experiments of drops impacting a smooth solid surface: a model of the critical impact speed for drop splashing},\ }\href@noop {} {\bibfield  {journal} {\bibinfo  {journal} {Phys. Rev. Lett.}\ }\textbf {\bibinfo {volume} {113}},\ \bibinfo {pages} {024507} (\bibinfo {year} {2014})}\BibitemShut {NoStop}%
\bibitem [{\citenamefont {Smith}\ \emph {et~al.}(2003)\citenamefont {Smith}, \citenamefont {Li},\ and\ \citenamefont {Wu}}]{smith:2003}%
  \BibitemOpen
  \bibfield  {author} {\bibinfo {author} {\bibfnamefont {F.}~\bibnamefont {Smith}}, \bibinfo {author} {\bibfnamefont {L.}~\bibnamefont {Li}},\ and\ \bibinfo {author} {\bibfnamefont {G.~X.}\ \bibnamefont {Wu}},\ }\bibfield  {title} {\bibinfo {title} {Air cushioning with a lubrication/inviscid balance},\ }\href {https://doi.org/10.1017/S0022112003004063} {\bibfield  {journal} {\bibinfo  {journal} {Journal of Fluid Mechanics}\ }\textbf {\bibinfo {volume} {482}},\ \bibinfo {pages} {291–318} (\bibinfo {year} {2003})}\BibitemShut {NoStop}%
\bibitem [{\citenamefont {Hicks}\ and\ \citenamefont {Purvis}(2010)}]{Hicks:2010}%
  \BibitemOpen
  \bibfield  {author} {\bibinfo {author} {\bibfnamefont {P.~D.}\ \bibnamefont {Hicks}}\ and\ \bibinfo {author} {\bibfnamefont {R.}~\bibnamefont {Purvis}},\ }\bibfield  {title} {\bibinfo {title} {Air cushioning and bubble entrapment in three-dimensional droplet impacts},\ }\href {https://doi.org/10.1017/S0022112009994009} {\bibfield  {journal} {\bibinfo  {journal} {Journal of Fluid Mechanics}\ }\textbf {\bibinfo {volume} {649}},\ \bibinfo {pages} {135–163} (\bibinfo {year} {2010})}\BibitemShut {NoStop}%
\bibitem [{\citenamefont {Hicks}\ \emph {et~al.}(2012)\citenamefont {Hicks}, \citenamefont {Ermanyuk}, \citenamefont {Gavrilov},\ and\ \citenamefont {Purvis}}]{Purvis:2012}%
  \BibitemOpen
  \bibfield  {author} {\bibinfo {author} {\bibfnamefont {P.~D.}\ \bibnamefont {Hicks}}, \bibinfo {author} {\bibfnamefont {E.~V.}\ \bibnamefont {Ermanyuk}}, \bibinfo {author} {\bibfnamefont {N.~V.}\ \bibnamefont {Gavrilov}},\ and\ \bibinfo {author} {\bibfnamefont {R.}~\bibnamefont {Purvis}},\ }\bibfield  {title} {\bibinfo {title} {Air trapping at impact of a rigid sphere onto a liquid},\ }\href {https://doi.org/10.1017/jfm.2012.20} {\bibfield  {journal} {\bibinfo  {journal} {Journal of Fluid Mechanics}\ }\textbf {\bibinfo {volume} {695}},\ \bibinfo {pages} {310–320} (\bibinfo {year} {2012})}\BibitemShut {NoStop}%
\bibitem [{\citenamefont {Hicks}\ and\ \citenamefont {Purvis}(2013)}]{hicks+purvis:2013}%
  \BibitemOpen
  \bibfield  {author} {\bibinfo {author} {\bibfnamefont {P.~D.}\ \bibnamefont {Hicks}}\ and\ \bibinfo {author} {\bibfnamefont {R.}~\bibnamefont {Purvis}},\ }\bibfield  {title} {\bibinfo {title} {Liquid–solid impacts with compressible gas cushioning},\ }\href {https://doi.org/10.1017/jfm.2013.487} {\bibfield  {journal} {\bibinfo  {journal} {Journal of Fluid Mechanics}\ }\textbf {\bibinfo {volume} {735}},\ \bibinfo {pages} {120–149} (\bibinfo {year} {2013})}\BibitemShut {NoStop}%
\bibitem [{\citenamefont {Wu}\ and\ \citenamefont {Bogy}(2001)}]{Lin:2001}%
  \BibitemOpen
  \bibfield  {author} {\bibinfo {author} {\bibfnamefont {L.}~\bibnamefont {Wu}}\ and\ \bibinfo {author} {\bibfnamefont {D.~B.}\ \bibnamefont {Bogy}},\ }\bibfield  {title} {\bibinfo {title} {A generalized compressible reynolds lubrication equation with bounded contact pressure},\ }\href {https://doi.org/10.1063/1.1384867} {\bibfield  {journal} {\bibinfo  {journal} {Physics of Fluids}\ }\textbf {\bibinfo {volume} {13}},\ \bibinfo {pages} {2237} (\bibinfo {year} {2001})}\BibitemShut {NoStop}%
\bibitem [{\citenamefont {Hillairet}(2007)}]{hillairet:2007}%
  \BibitemOpen
  \bibfield  {author} {\bibinfo {author} {\bibfnamefont {M.}~\bibnamefont {Hillairet}},\ }\bibfield  {title} {\bibinfo {title} {Lack of collision between solid bodies in a 2d incompressible viscous flow},\ }\href@noop {} {\bibfield  {journal} {\bibinfo  {journal} {Communications in Partial Differential Equations}\ }\textbf {\bibinfo {volume} {32}},\ \bibinfo {pages} {1345} (\bibinfo {year} {2007})}\BibitemShut {NoStop}%
\bibitem [{\citenamefont {Langley}\ \emph {et~al.}(2017)\citenamefont {Langley}, \citenamefont {Li},\ and\ \citenamefont {Thoroddsen}}]{langley:2017}%
  \BibitemOpen
  \bibfield  {author} {\bibinfo {author} {\bibfnamefont {K.~R.}\ \bibnamefont {Langley}}, \bibinfo {author} {\bibfnamefont {E.~Q.}\ \bibnamefont {Li}},\ and\ \bibinfo {author} {\bibfnamefont {S.~T.}\ \bibnamefont {Thoroddsen}},\ }\bibfield  {title} {\bibinfo {title} {Impact of ultra-viscous drops: air-film gliding and extreme wetting},\ }\href@noop {} {\bibfield  {journal} {\bibinfo  {journal} {Journal of Fluid Mechanics}\ }\textbf {\bibinfo {volume} {813}},\ \bibinfo {pages} {647} (\bibinfo {year} {2017})}\BibitemShut {NoStop}%
\bibitem [{\citenamefont {Langley}\ and\ \citenamefont {Thoroddsen}(2019)}]{langley:2019}%
  \BibitemOpen
  \bibfield  {author} {\bibinfo {author} {\bibfnamefont {K.~R.}\ \bibnamefont {Langley}}\ and\ \bibinfo {author} {\bibfnamefont {S.~T.}\ \bibnamefont {Thoroddsen}},\ }\bibfield  {title} {\bibinfo {title} {Gliding on a layer of air: impact of a large-viscosity drop on a liquid film},\ }\href@noop {} {\bibfield  {journal} {\bibinfo  {journal} {Journal of Fluid Mechanics}\ }\textbf {\bibinfo {volume} {878}},\ \bibinfo {pages} {R2} (\bibinfo {year} {2019})}\BibitemShut {NoStop}%
\bibitem [{\citenamefont {Langley}\ \emph {et~al.}(2020)\citenamefont {Langley}, \citenamefont {Castrej{\'o}n-Pita},\ and\ \citenamefont {Thoroddsen}}]{langley:2020}%
  \BibitemOpen
  \bibfield  {author} {\bibinfo {author} {\bibfnamefont {K.~R.}\ \bibnamefont {Langley}}, \bibinfo {author} {\bibfnamefont {A.~A.}\ \bibnamefont {Castrej{\'o}n-Pita}},\ and\ \bibinfo {author} {\bibfnamefont {S.~T.}\ \bibnamefont {Thoroddsen}},\ }\bibfield  {title} {\bibinfo {title} {Droplet impacts onto soft solids entrap more air},\ }\href@noop {} {\bibfield  {journal} {\bibinfo  {journal} {Soft Matter}\ }\textbf {\bibinfo {volume} {16}},\ \bibinfo {pages} {5702} (\bibinfo {year} {2020})}\BibitemShut {NoStop}%
\bibitem [{\citenamefont {Bouwhuis}\ \emph {et~al.}(2012)\citenamefont {Bouwhuis}, \citenamefont {van~der Veen}, \citenamefont {Tran}, \citenamefont {Keij}, \citenamefont {Winkels}, \citenamefont {Peters}, \citenamefont {van~der Meer}, \citenamefont {Sun}, \citenamefont {Snoeijer},\ and\ \citenamefont {Lohse}}]{Bouwhuis:2012}%
  \BibitemOpen
  \bibfield  {author} {\bibinfo {author} {\bibfnamefont {W.}~\bibnamefont {Bouwhuis}}, \bibinfo {author} {\bibfnamefont {R.~C.~A.}\ \bibnamefont {van~der Veen}}, \bibinfo {author} {\bibfnamefont {T.}~\bibnamefont {Tran}}, \bibinfo {author} {\bibfnamefont {D.~L.}\ \bibnamefont {Keij}}, \bibinfo {author} {\bibfnamefont {K.~G.}\ \bibnamefont {Winkels}}, \bibinfo {author} {\bibfnamefont {I.~R.}\ \bibnamefont {Peters}}, \bibinfo {author} {\bibfnamefont {D.}~\bibnamefont {van~der Meer}}, \bibinfo {author} {\bibfnamefont {C.}~\bibnamefont {Sun}}, \bibinfo {author} {\bibfnamefont {J.~H.}\ \bibnamefont {Snoeijer}},\ and\ \bibinfo {author} {\bibfnamefont {D.}~\bibnamefont {Lohse}},\ }\bibfield  {title} {\bibinfo {title} {Maximal air bubble entrainment at liquid-drop impact},\ }\href {https://doi.org/10.1103/PhysRevLett.109.264501} {\bibfield  {journal} {\bibinfo  {journal} {Phys. Rev. Lett.}\ }\textbf {\bibinfo {volume} {109}},\ \bibinfo {pages} {264501} (\bibinfo {year} {2012})}\BibitemShut {NoStop}%
\bibitem [{\citenamefont {Balmforth}\ \emph {et~al.}(2010)\citenamefont {Balmforth}, \citenamefont {Cawthorn},\ and\ \citenamefont {R.V.}}]{balmforth:2010}%
  \BibitemOpen
  \bibfield  {author} {\bibinfo {author} {\bibfnamefont {N.}~\bibnamefont {Balmforth}}, \bibinfo {author} {\bibfnamefont {C.}~\bibnamefont {Cawthorn}},\ and\ \bibinfo {author} {\bibfnamefont {C.}~\bibnamefont {R.V.}},\ }\bibfield  {title} {\bibinfo {title} {Contact in a viscous fluid. part 2. a compressible fluid and an elastic solid},\ }\href {https://doi.org/10.1017/S0022112009993168} {\bibfield  {journal} {\bibinfo  {journal} {Journal of Fluid Mechanics}\ }\textbf {\bibinfo {volume} {646}},\ \bibinfo {pages} {339–361} (\bibinfo {year} {2010})}\BibitemShut {NoStop}%
\bibitem [{\citenamefont {Davis}\ \emph {et~al.}(1986)\citenamefont {Davis}, \citenamefont {Serayssol},\ and\ \citenamefont {Hinch}}]{David:1986}%
  \BibitemOpen
  \bibfield  {author} {\bibinfo {author} {\bibfnamefont {R.~H.}\ \bibnamefont {Davis}}, \bibinfo {author} {\bibfnamefont {J.-M.}\ \bibnamefont {Serayssol}},\ and\ \bibinfo {author} {\bibfnamefont {E.~J.}\ \bibnamefont {Hinch}},\ }\bibfield  {title} {\bibinfo {title} {The elastohydrodynamic collision of two spheres},\ }\href {https://doi.org/10.1017/S0022112086002392} {\bibfield  {journal} {\bibinfo  {journal} {Journal of Fluid Mechanics}\ }\textbf {\bibinfo {volume} {163}},\ \bibinfo {pages} {479–497} (\bibinfo {year} {1986})}\BibitemShut {NoStop}%
\bibitem [{\citenamefont {Duchemin}\ and\ \citenamefont {Josserand}(2011)}]{Duchemin-Josserand:2011}%
  \BibitemOpen
  \bibfield  {author} {\bibinfo {author} {\bibfnamefont {L.}~\bibnamefont {Duchemin}}\ and\ \bibinfo {author} {\bibfnamefont {C.}~\bibnamefont {Josserand}},\ }\bibfield  {title} {\bibinfo {title} {{Curvature singularity and film-skating during drop impact}},\ }\href {https://doi.org/10.1063/1.3640028} {\bibfield  {journal} {\bibinfo  {journal} {Physics of Fluids}\ }\textbf {\bibinfo {volume} {23}},\ \bibinfo {pages} {091701} (\bibinfo {year} {2011})},\ \Eprint {https://arxiv.org/abs/https://pubs.aip.org/aip/pof/article-pdf/doi/10.1063/1.3640028/15888818/091701\_1\_online.pdf} {https://pubs.aip.org/aip/pof/article-pdf/doi/10.1063/1.3640028/15888818/091701\_1\_online.pdf} \BibitemShut {NoStop}%
\bibitem [{\citenamefont {Gordillo}\ and\ \citenamefont {Riboux}(2022)}]{Gordillo:2022}%
  \BibitemOpen
  \bibfield  {author} {\bibinfo {author} {\bibfnamefont {J.~M.}\ \bibnamefont {Gordillo}}\ and\ \bibinfo {author} {\bibfnamefont {G.}~\bibnamefont {Riboux}},\ }\bibfield  {title} {\bibinfo {title} {The initial impact of drops cushioned by an air or vapour layer with applications to the dynamic leidenfrost regime},\ }\href {https://doi.org/10.1017/jfm.2022.280} {\bibfield  {journal} {\bibinfo  {journal} {Journal of Fluid Mechanics}\ }\textbf {\bibinfo {volume} {941}},\ \bibinfo {pages} {A10} (\bibinfo {year} {2022})}\BibitemShut {NoStop}%
\bibitem [{\citenamefont {Sprittles}(2024)}]{Sprittles:2023}%
  \BibitemOpen
  \bibfield  {author} {\bibinfo {author} {\bibfnamefont {J.~E.}\ \bibnamefont {Sprittles}},\ }\bibfield  {title} {\bibinfo {title} {Gas microfilms in droplet dynamics: When do drops bounce?},\ }\bibfield  {journal} {\bibinfo  {journal} {Annual Review of Fluid Mechanics}\ }\textbf {\bibinfo {volume} {56}},\ \href {https://doi.org/10.1146/annurev-fluid-121021-021121} {10.1146/annurev-fluid-121021-021121} (\bibinfo {year} {2024}),\ \Eprint {https://arxiv.org/abs/https://doi.org/10.1146/annurev-fluid-121021-021121} {https://doi.org/10.1146/annurev-fluid-121021-021121} \BibitemShut {NoStop}%
\bibitem [{\citenamefont {de~Ruiter}\ \emph {et~al.}(2012)\citenamefont {de~Ruiter}, \citenamefont {Oh}, \citenamefont {van~den Ende},\ and\ \citenamefont {Mugele}}]{Ruiter:2012}%
  \BibitemOpen
  \bibfield  {author} {\bibinfo {author} {\bibfnamefont {J.}~\bibnamefont {de~Ruiter}}, \bibinfo {author} {\bibfnamefont {J.~M.}\ \bibnamefont {Oh}}, \bibinfo {author} {\bibfnamefont {D.}~\bibnamefont {van~den Ende}},\ and\ \bibinfo {author} {\bibfnamefont {F.}~\bibnamefont {Mugele}},\ }\bibfield  {title} {\bibinfo {title} {Dynamics of collapse of air films in drop impact},\ }\href {https://doi.org/10.1103/PhysRevLett.108.074505} {\bibfield  {journal} {\bibinfo  {journal} {Phys. Rev. Lett.}\ }\textbf {\bibinfo {volume} {108}},\ \bibinfo {pages} {074505} (\bibinfo {year} {2012})}\BibitemShut {NoStop}%
\bibitem [{\citenamefont {Szeri}(2010)}]{szeri:2010}%
  \BibitemOpen
  \bibfield  {author} {\bibinfo {author} {\bibfnamefont {A.~Z.}\ \bibnamefont {Szeri}},\ }\href {https://doi.org/10.1017/CBO9780511782022} {\emph {\bibinfo {title} {Fluid Film Lubrication}}},\ \bibinfo {edition} {2nd}\ ed.\ (\bibinfo  {publisher} {Cambridge University Press},\ \bibinfo {year} {2010})\BibitemShut {NoStop}%
\bibitem [{\citenamefont {Savitski}\ and\ \citenamefont {Detournay}(2002)}]{savitski:2002}%
  \BibitemOpen
  \bibfield  {author} {\bibinfo {author} {\bibfnamefont {A.~A.}\ \bibnamefont {Savitski}}\ and\ \bibinfo {author} {\bibfnamefont {E.}~\bibnamefont {Detournay}},\ }\bibfield  {title} {\bibinfo {title} {Propagation of a penny-shaped fluid-driven fracture in an impermeable rock: asymptotic solutions},\ }\href@noop {} {\bibfield  {journal} {\bibinfo  {journal} {International journal of solids and structures}\ }\textbf {\bibinfo {volume} {39}},\ \bibinfo {pages} {6311} (\bibinfo {year} {2002})}\BibitemShut {NoStop}%
\bibitem [{\citenamefont {Sautter}\ \emph {et~al.}(2022)\citenamefont {Sautter}, \citenamefont {Me{\ss}mer}, \citenamefont {Teschemacher},\ and\ \citenamefont {Bletzinger}}]{sautter:2022}%
  \BibitemOpen
  \bibfield  {author} {\bibinfo {author} {\bibfnamefont {K.~B.}\ \bibnamefont {Sautter}}, \bibinfo {author} {\bibfnamefont {M.}~\bibnamefont {Me{\ss}mer}}, \bibinfo {author} {\bibfnamefont {T.}~\bibnamefont {Teschemacher}},\ and\ \bibinfo {author} {\bibfnamefont {K.-U.}\ \bibnamefont {Bletzinger}},\ }\bibfield  {title} {\bibinfo {title} {Limitations of the st. venant--kirchhoff material model in large strain regimes},\ }\href@noop {} {\bibfield  {journal} {\bibinfo  {journal} {International Journal of Non-Linear Mechanics}\ }\textbf {\bibinfo {volume} {147}},\ \bibinfo {pages} {104207} (\bibinfo {year} {2022})}\BibitemShut {NoStop}%
\bibitem [{\citenamefont {Cimbala}\ and\ \citenamefont {Cengel}(2006)}]{cimbala:2006}%
  \BibitemOpen
  \bibfield  {author} {\bibinfo {author} {\bibfnamefont {J.~M.}\ \bibnamefont {Cimbala}}\ and\ \bibinfo {author} {\bibfnamefont {Y.~A.}\ \bibnamefont {Cengel}},\ }\href@noop {} {\emph {\bibinfo {title} {Fluid mechanics: fundamentals and applications}}}\ (\bibinfo  {publisher} {McGraw-Hill Higher Education},\ \bibinfo {year} {2006})\BibitemShut {NoStop}%
\bibitem [{\citenamefont {Brooks}\ and\ \citenamefont {Hughes}(1982)}]{Brooks:1982}%
  \BibitemOpen
  \bibfield  {author} {\bibinfo {author} {\bibfnamefont {A.~N.}\ \bibnamefont {Brooks}}\ and\ \bibinfo {author} {\bibfnamefont {T.~J.~R.}\ \bibnamefont {Hughes}},\ }\bibfield  {title} {\bibinfo {title} {Streamline upwind/petrov-galerkin formulations for convection dominated flows with particular emphasis on the incompressible navier-stokes equations},\ }\href {https://doi.org/https://doi.org/10.1016/0045-7825(82)90071-8} {\bibfield  {journal} {\bibinfo  {journal} {Computer Methods in Applied Mechanics and Engineering}\ }\textbf {\bibinfo {volume} {32}},\ \bibinfo {pages} {199} (\bibinfo {year} {1982})}\BibitemShut {NoStop}%
\bibitem [{\citenamefont {Habchi}\ \emph {et~al.}(2008)\citenamefont {Habchi}, \citenamefont {Eyheramendy}, \citenamefont {Vergne},\ and\ \citenamefont {Morales-Espejel}}]{habchi:2008}%
  \BibitemOpen
  \bibfield  {author} {\bibinfo {author} {\bibfnamefont {W.}~\bibnamefont {Habchi}}, \bibinfo {author} {\bibfnamefont {D.}~\bibnamefont {Eyheramendy}}, \bibinfo {author} {\bibfnamefont {P.}~\bibnamefont {Vergne}},\ and\ \bibinfo {author} {\bibfnamefont {G.}~\bibnamefont {Morales-Espejel}},\ }\bibfield  {title} {\bibinfo {title} {A full-system approach of the elastohydrodynamic line-point contact problem},\ }\href@noop {} {\bibfield  {journal} {\bibinfo  {journal} {Journal of Tribology}\ } (\bibinfo {year} {2008})}\BibitemShut {NoStop}%
\bibitem [{Note1()}]{Note1}%
  \BibitemOpen
  \bibinfo {note} {We used the inertial regime scales to adimensionalize the transition, but using the elastic regime's ones would yield similar results}\BibitemShut {NoStop}%
\bibitem [{\citenamefont {Kolinski}\ \emph {et~al.}(2014)\citenamefont {Kolinski}, \citenamefont {Mahadevan},\ and\ \citenamefont {Rubinstein}}]{kolinski:2014}%
  \BibitemOpen
  \bibfield  {author} {\bibinfo {author} {\bibfnamefont {J.~M.}\ \bibnamefont {Kolinski}}, \bibinfo {author} {\bibfnamefont {L.}~\bibnamefont {Mahadevan}},\ and\ \bibinfo {author} {\bibfnamefont {S.~M.}\ \bibnamefont {Rubinstein}},\ }\bibfield  {title} {\bibinfo {title} {Lift-off instability during the impact of a drop on a solid surface},\ }\href {https://doi.org/10.1103/PhysRevLett.112.134501} {\bibfield  {journal} {\bibinfo  {journal} {Phys. Rev. Lett.}\ }\textbf {\bibinfo {volume} {112}},\ \bibinfo {pages} {134501} (\bibinfo {year} {2014})}\BibitemShut {NoStop}%
\bibitem [{\citenamefont {Batchelor}(2000)}]{batchelor:2000}%
  \BibitemOpen
  \bibfield  {author} {\bibinfo {author} {\bibfnamefont {G.~K.}\ \bibnamefont {Batchelor}},\ }\href {https://doi.org/10.1017/CBO9780511800955} {\emph {\bibinfo {title} {An Introduction to Fluid Dynamics}}},\ Cambridge Mathematical Library\ (\bibinfo  {publisher} {Cambridge University Press},\ \bibinfo {year} {2000})\BibitemShut {NoStop}%
\bibitem [{\citenamefont {Landau}\ and\ \citenamefont {Lifshitz}(2013)}]{landau:2013}%
  \BibitemOpen
  \bibfield  {author} {\bibinfo {author} {\bibfnamefont {L.~D.}\ \bibnamefont {Landau}}\ and\ \bibinfo {author} {\bibfnamefont {E.~M.}\ \bibnamefont {Lifshitz}},\ }\href@noop {} {\emph {\bibinfo {title} {Fluid Mechanics: Landau and Lifshitz: Course of Theoretical Physics, Volume 6}}},\ Vol.~\bibinfo {volume} {6}\ (\bibinfo  {publisher} {Elsevier},\ \bibinfo {year} {2013})\BibitemShut {NoStop}%
\bibitem [{\citenamefont {Li}\ \emph {et~al.}(2019)\citenamefont {Li}, \citenamefont {Wang},\ and\ \citenamefont {Dong}}]{li:2019}%
  \BibitemOpen
  \bibfield  {author} {\bibinfo {author} {\bibfnamefont {P.~F.}\ \bibnamefont {Li}}, \bibinfo {author} {\bibfnamefont {S.~F.}\ \bibnamefont {Wang}},\ and\ \bibinfo {author} {\bibfnamefont {W.~L.}\ \bibnamefont {Dong}},\ }\bibfield  {title} {\bibinfo {title} {Capillary wave and initial spreading velocity at impact of drop onto a surface},\ }\href@noop {} {\bibfield  {journal} {\bibinfo  {journal} {Journal of Applied Fluid Mechanics}\ }\textbf {\bibinfo {volume} {12}},\ \bibinfo {pages} {1265} (\bibinfo {year} {2019})}\BibitemShut {NoStop}%
\bibitem [{\citenamefont {Marsden}\ and\ \citenamefont {Hughes}(1994)}]{Mardsen:1994}%
  \BibitemOpen
  \bibfield  {author} {\bibinfo {author} {\bibfnamefont {J.~E.}\ \bibnamefont {Marsden}}\ and\ \bibinfo {author} {\bibfnamefont {T.~J.~R.}\ \bibnamefont {Hughes}},\ }\href@noop {} {\emph {\bibinfo {title} {Mathematical foundations of elasticity}}}\ (\bibinfo  {publisher} {Courier Corporation},\ \bibinfo {year} {1994})\BibitemShut {NoStop}%
\end{thebibliography}%

\end{document}